# Isogenies of Elliptic Curves: A Computational Approach

Daniel Shumow

August 13, 2009

# Part I
# Introduction

The study of elliptic curves has historically been a subject of almost purely mathematical interest. However, Koblitz and Miller independently showed that elliptic curves can be used to implement cryptographic primitives [13], [17]. This thrust elliptic curves from the abstract realm of pure mathematics to the preeminently applied world of communications security. Public key cryptography in general advanced the development of the Internet and was in turn further advanced by this new use. Elliptic curve cryptography (ECC) was also developed and advanced along with the general field of public key cryptography.

Elliptic curves provide benefits over the groups previously proposed for use in cryptography. Unlike finite fields, elliptic curves do not have a ring structure (the two related group operations of addition and multiplication), and hence are not vulnerable to index calculus like attacks [12]. The direct effect of this is that using elliptic curves over smaller finite fields yields the same security as using discrete log or factoring based public key crypto systems of Diffie-Hellman and RSA with larger moduli. This makes ECC ideally suited to small embedded and low power devices such as cell phones. So it is unsurprising that as these type of small devices have increased in popularity in recent years, ECC has as well.

As elliptic curves are now used in cryptography, the computational aspects of them have real world applications. The underlying theory is very deep and touches on many different branches of mathematics. Elliptic curves have a very rich mathematical structure and the subject of ECC is about determining how to best apply and efficiently compute with this deep structure.

The maps defined on any mathematical object are a key part of the underlying structure. In the case of elliptic curves, the principal maps of interest are the isogenies. An isogeny is a non-constant function, defined on an elliptic curve, that takes values on another elliptic curve and preserves point addition. In short, isogenies are functions that preserve the elliptic curve structure. As such, they are a powerful tool for studying elliptic curves and similar to elliptic curves admit a deep underlying theory that is interesting from many different perspectives such as complex analysis, algebra, number theory, and algebraic geometry.

In addition to providing an abstract tool for the study of elliptic curves, isogenies are concrete mathematical objects that can be written down and used for computations. Vélu's formulas [22] initially provided an algorithm to compute the codomain and rational maps given the domain and kernel of an isogeny. This work has been greatly expanded upon and improved by subsequent authors [16], [2]. Furthermore, the problem of computing an isogeny given the domain and codomain is also well understood.

With the advent of elliptic curve cryptography, isogenies have found an application in cryptology as well. These applications provide motivation for a more widespread audience to understand and use them. Here we provide a brief list of these uses and their relevance to the greater field.

The first application of isogenies to cryptography was as a tool in the Schoof-Elkies-Atkins (SEA) algorithm for counting the number of points on elliptic curves over finite fields [1]. Originally Schoof had provided an algorithm that, when given a curve $E$ defined over some finite field $\mathbb{F}_q$, would return the number of points in the group of points on $E$ defined over $\mathbb{F}_q$. Schoof's original algorithm was polynomial time, but if $n$ is the number of bits in $q$ then by using straight forward arithmetic has a complexity of $O(n^{5+\epsilon})$. The SEA improvement results in a complexity of $O(n^{4+\epsilon})$ which is significant at such degrees. This improvement fundamentally uses isogenies.

More recently, isogenies have been used as a tool to analyze the computational difficulty of the elliptic curve discrete log problem (ECDLP) [9]. Specifically, the paper shows that isogenies can be used to create a randomized algorithm that will reduce the ECDLP from one set of curves to a significantly larger set of curves in polynomial time. The authors argue that this provides complexity theoretic evidence that the difficulty of discrete logs on all curves of the same order is the same.

Isogenies have also been proposed as a tool in constructing random number generators and hash functions [6]. In particular, isogenies can be used as a one way function that can be used in these cryptographic primitives. The nice mathematical properties lend themselves to a rigorous analysis of the security properties. In turn, these hash functions and random number generators can be considered provably secure with respect to some hardness assumptions.

Before the introduction of elliptic curves to cryptography, few people in the field of computer security would be worried about the most efficient way to implement elliptic curve arithmetic. However, this is now a deep and popular area of research. As isogenies are a tool used in cryptography there is a need for the field to be more accessible to people without a deep mathematical background.

This document includes an introduction to the basic theory of isogenies of elliptic curves, viewing them as a generalization of the multiplication by $m$ map. This is presented in a fashion that only presupposes a familiarity of elliptic curves and abstract algebra at the level one would need to be comfortable with the subject of elliptic curve cryptography. After an introduction to the basic theory, there are several algorithms for computational aspects of isogenies. These algorithms focus on how to represent isogenies, and how to deduce one representation from another. For example, one such method is to determine the codomain and coordinate maps of an isogeny from the kernel. Another method determines the kernel and rational maps from the domain and codomain. The algorithms are presented with proofs of correctness, as well as analyses of the computational complexity.

# Part II
# Basic Theory

Throughout this section, unless otherwise noted, we will use the following notation:

$K$ - A field.
$\overline{K}$ - A fixed algebraic closure of K.
$E$ - A fixed elliptic curve given by the Weierstrass model

$$y^2 + a_1xy + a_3y = x^3 + a_2x^2 + a_4x + a_6$$

with coefficients in $K$.
$E(K), E(\overline{K})$ - The set of pairs $(x, y)$ satisfying the Weierstrass equation of $E$ where $x$ and $y$ are taken in $K$ or $\overline{K}$ respectively.
$\varphi$ - An isogeny (to be defined later) from $E$ to another elliptic curve $E'$.

Also, in this exposition, we have tried to give and prove the most general results. For ease of reading and understanding, most textbook presentations such as Silverman [20] and Washington [23] both tend to assume that the curve $E$ is in short Weierstrass form, and assume that the characteristic of $K$ is not 2 or 3. Whenever possible we have favored results that work for a curve in general Weierstrass form, and try to avoid making conditions on the characteristic of $K$ as much as is reasonable. For the most part, this does not significantly affect the proofs or reasoning, aside from adding technical details, that admittedly makes them a little bit more messy. However, the results for curves in short Weierstrass form, as well as the characteristic 2, or 3 case, follow immediately from the general results. Also, the hope is that for a reader who is not as familiar with the techniques being used, can consult these proofs if they do not see how to generalize the results in the canonical introductory texts.

## 1    The Multiplication By $m$ Map

We are interested in studying maps that preserve both the group structure, and the structure of an elliptic curve as an algebraic variety. It is instructive to see if there are any such maps immediately at our disposal. Indeed, one such map is the multiplication by $m$ map, This map is the usual map computed by adding a point to itself $m$-times, and is very familiar in elliptic curve cryptography, as it is the principal operation in ECDH and ECDSA [23]. It is clear that this map preserves point addition, and it maps the curve $E$ to itself. As this map satisfies the properties we are interested in, we shall now investigate it in some detail.

Recall that the elliptic curve $E$ group operation "point addition" ([20] section III.2) is such that for points $P = (x_P, y_P)$ and $Q = (x_Q, y_Q)$ on $E$ the sum is given by the formulas:

$$(x_{P+Q}, y_{P+Q}) = (x_P, y_P) + (x_Q, y_Q).$$

The formulas for $x_{P+Q}$ and $y_{P+Q}$ are

$$x_{P+Q} = \lambda^2 + a_1 \lambda - a_2 - x_P - x_Q$$

and

$$y_{P+Q} = -(\lambda + a_1) x_{P+Q} - \nu - a_3$$

where $\lambda$ and $\nu$ are given as follows. If $x_P \neq x_Q$ then

$$\lambda = \frac{y_p - y_Q}{x_P - x_Q} \text{ and } \nu = \frac{y_Q x_P - y_P x_Q}{x_P - x_Q},$$

and otherwise when $x_P = x_Q$ then

$$\lambda = \frac{3x_P^2 + 2a_2 x_P + a_4 - a_1 y_P}{2 y_P + a_1 x_P + a_3} \text{ and } \nu = \frac{-x_P^3 + a_4 x_P + 2a_6 - a_3 y_P}{2 y_P + a_1 x_P + a_3}.$$

Furthermore if $P = (x, y)$ then

$$-P = (x, -y - a_1 x - a_3).$$

Immediately, this gives the duplication formula:

$$x_{2P} = \frac{x^4 - b_4 x^2 - 2b_6 x - b_8}{4x^3 + b_2 x^2 + 2b_4 x + b_6}, \tag{1}$$

where $b_2$, $b_4$, $b_6$ and $b_8$ are the $b$-invariants given in [20] section III.1. This formula can be substituted in to derive a similar formula for $y_{2P}$.

So this gives a nice rational map for multiplication by 2 on a curve $E$. Thus using a "double-and-add" approach with this formula and the addition formulas, If $P = (x, y)$ denotes a non-infinite point on $E$, then there are rational maps for the coordinates of $mP$. The rest of this section is devoted to generalizing the duplication formula to give clear formulas for multiplication by $m$.

First note that if a point $P$ on $E$ is a two torsion point, meaning $2P$ is the point at infinity, denoted here as $\infty$, then $P = -P$. So

$$y_P = -y_P - a_1 x_P - a_3$$

therefore
$$2y_P + a_1 x_P + a_3 = 0. \tag{2}$$
So any two torsion points on $E$ must satisfy this equation. However, we also have the fact that if $P$ is a two torsion point then the duplication formula for $x_{2P}$ must go to infinity (because the $x$-coordinate function on $E$ has a pole at infinity.) This implies that the denominator of the rational map in (2.1) evaluates to 0. Therefore if $P = (x, y)$ is a two torsion point then
$$4x^3 + b_2 x^2 + 2b_4 x + b_6 = 0. \tag{3}$$
Which can be seen quite precisely, when working in characteristic not equal to 2 and replacing $y$ by $\frac{1}{2}(y - a_1 x - a_3)$ in the Weierstrass equation for $E$ and solving for $y^2$ gives
$$y^2 = 4x^3 + b_2 x^2 + 2b_4 x + b_6$$
and the points with $y = 0$ in this new equation are the points satisfying equation (2.3) on $E$. Thus the values satisfying (2.3) are $x$ coordinates of a two torsion point on $E$. Herein, we refer to the polynomial in equation (2.2) as the bivariate two torsion polynomial, and the polynomial in equation (2.3) as the univariate two torsion polynomial.

For the rest of the discussion of the multiplication by $m$ map, to simplify the presentation we diverge from the general approach and restrict our attention to the case of characteristic not 2 or 3. So we can assume that our curves are in short Weierstrass form:
$$y^2 = x^3 + Ax + B.$$
The treatment in this section follows [20] exercise 3.7 and [23] section 3.2. The reader interested in the values for the general case can see [1] section III.4.

Immediately from the assumption that the curves are in short Weierstrass form we get $a_1 = a_3 = 0$, this gives that the bivariate two torsion polynomial is $2y$.

**Definition 1.1.** The *torsion polynomials* are polynomials in $\mathbb{Z}[A, B, x, y, (2y)^{-1}]$. The first four are defined explicitly as
$$\begin{aligned}
\psi_1 &= 1 \\
\psi_2 &= 2y \\
\psi_3 &= 3x^4 + 6Ax^2 + 12Bx - A^2 \\
\psi_4 &= 4y(x^6 + 5Ax^4 + 20Bx^3 - 5A^2 x^2 - 4ABx - 8B^2 - A^3)
\end{aligned}$$

the subsequent polynomials are defined inductively as
$$\begin{aligned}
\psi_{2m+1} &= \psi_{m+2}\psi_m^3 - \psi_{m-1}\psi_{m+1}^3 & (m \geq 2) \\
\psi_{2m} &= (2y)^{-1}\psi_m(\psi_{m+2}\psi_{m-1}^2 - \psi_{m-2}\psi_{m+1}^2) & (m \geq 3)
\end{aligned}$$

*Remark* 1.2. Some authors prefer the term *division polynomial* to *torsion polynomial*, however herein they mean exactly the same thing.

*Remark* 1.3. The torsion polynomials $\psi_3$ and $\psi_4$ are found by looking for the polynomials that evaluate to 0, when $P = (x, y)$ is a 3 or 4 torsion point, similar to the approach for finding the 2-torsion polynomial.

It is somewhat awkward to have the torsion polynomials be defined over some fraction involving the variable $y$ in the denominator. However, it turns out that we can take the torsion polynomials to be in a less awkwardly defined polynomial ring.

**Lemma 1.4.** *For all positive integers $m$ the division polynomial $\psi_m$ is contained in the polynomial ring $\mathbb{Z}[A, B, x, y]$.*

*Proof.* Just by looking at the formulas, we can see that the confusion only comes in on the definition of $\psi_{2m}$ as that is the only definition that includes the denominator $2y$.

We prove this lemma by arguing that not only is $\psi_{2m}$ a polynomial, it is evenly divisible by $2y$. Clearly this holds when $m$ is 1 and 2. For $m$ greater than 3 assume that the hypothesis holds for $n$ up to (but not including) $2m$ Now suppose that $m$, then $m - 2$, $m$, and $m + 2$ are divisible by $2y$, factoring these out of the recurrence shows that the denominator cancels one and the resulting expression is a polynomial still evenly divisible by $2y$. In the case that $m$ is odd then $m - 1$ and $m + 1$ are even, so by the induction hypothesis $\psi_{m-1}$ and $\psi_{m+1}$ are polynomials divisible by $2y$. Thus substituting this into the recurrence relation shows that the numerator is divisible by $(2y)^2$. The denominator cancels out one factor of $2y$ leaving the result as a polynomial that is divisible by $2y$. □

Finally given the torsion polynomials we define the polynomials

$$\phi_m = x\psi_m^2 - \psi_{m+1}\psi_{m-1}$$
$$\omega_m = \psi_{m+2}\psi_{m-1}^2 - \psi_{m-2}\psi_{m+1}^2.$$

These polynomials arise in the multiplication by $m$ map that we are trying to derive.

Next we make and prove some statements about the form of these polynomials.

**Lemma 1.5.** *When $m$ is odd $\psi_m$, $\phi_m$ and $y^{-1}\omega_m$ are polynomials in $\mathbb{Z}[A, B, x, y^2]$. When $m$ is even $((2y)^{-1}\psi_m, \phi_m,$ and $\omega_m$ are polynomials in $\mathbb{Z}[A, B, x, y^2]$.*

*Proof.* For simplicity, in this proof, we will let $R$ be $\mathbb{Z}[A, B, x, y^2]$. We will prove the cases of $\psi_m$, $\phi_m$ and $\omega_m$ separately.

First we show this is true for $\psi_m$. For $m \leq 4$, this can be seen by observation of the given formulas. Next we assume that the properties hold for $m < 2n$, where $2 < n$, so that $n + 2 < 2n$. Hence the inductive hypothesis holds for all of the formulas in the recurrence relation. In the case that $n$ is even then $n - 2$ and $n + 2$ are even as well, then $(2y)^{-1}\psi_i$ is a polynomial in $R$ for $i = n - 2, n, n + 2$. Also, $n - 1$, $n + 1$ are odd so $\psi_{n+1}$ and $\psi_{n-1}$ are in $R$. Hence plugging these values into the recurrence relation shows that $(2y)^{-1}\psi_{2n}$ is in $R$. In the case that $n$ is odd, then $(2y)^{-1}\psi_{n-1}$ and $(2y)^{-1}\psi_{m+1}$ are in $R$ as $m-1$ and $m+1$ are even. And $m+2$, $m$, and $m-2$ are odd so $\psi_{m+2}$, $\psi_m$ and $\psi_{m-2}$ are in $R$. Thus plugging into the recurrence relation gives that $(2y)^{-1}\psi_{2m}$ is in $R$.

Now we show that the lemma holds for $\psi_{2m+1}$. We assume that the properties hold for $m < 2n$, where $2 \leq n$, so that $n + 2 < 2n + 1$. If $n$ is even then $\psi_{n+2}$ and $\psi_n$ are in $2yR$ so $\psi_{n+2}\psi_n^3$ is in $(2y)^4 R$ which is contained in R (by replacing $y^2$ by $x^3 + Ax + B$). Also, $n-1$ and $n+1$ are odd so that $\psi_{n-1}\psi_{n+1}^3$ are in $R$, hence $\psi_{2n+1}$ is in R, by the recurrence relation. The case for $n$ odd is exactly symmetric (the even and odd values are reversed) and it follows that $\psi_{2m+1}$ is in $R$.

Next we examine $\phi_m$. If m is even then $\psi_m$ is in $2yR$ and hence $\psi_m^2$ is in $(2y)^2 R$ which is contained in $R$. Furthermore, $m-1$ and $m+1$ are odd, so $\psi_{m-1}$ and $\psi_{m+1}$ are in $R$. Thus $\phi_m$ is in R. If $m$ is odd, then $\psi_m$ is in $R$ and $\psi_{m-1}\psi_{m+1}$ are in $(2y)^2 R$ hence in $\mathbb{Z}[A, B, x, y^2]$. So $\phi_m$ is in $R$.

Finally, we show that the properties hold for $\omega_m$. If $m$ is odd then so are $m+2$ and $m-2$, so that $\psi_{m+2}$ and $\psi_{m-2}$ are in $R$. Also, $m+1$ and $m-1$ are even so that $\psi_{m+1}$ and $\psi_{m-1}$ are in $2yR$ so $\psi_{m+1}^2$ and $\psi_{m-1}^2$ are in $(2y)^2R$. Thus by the definition of $\omega_n$ it is in $R$. If $m$ is even then $(2y)^{-1}\psi_{m+2}$ and $(2y)^{-1}\psi_{m-2}$ are in $R$ and $\psi_{m-1}$ and $\psi_{m+1}$ are in $R$ as well. Hence

$$2\omega_m = (2y)^{-1}\psi_{m+2}\psi_{m-1}^2 - (2y)^{-1}\psi_{m-2}\psi_{m+1}^2 \tag{4}$$

is in $R$. Now we prove the following by induction: When $m$ is odd

$$\psi_m \equiv (x^2 + A)^{(m^2-1)/4} \mod 2 \tag{5}$$

and when $m$ is even

$$(2y)^{-1}\psi_m \equiv (m/2)(x^2 + A)^{(m^2-4)/4} \mod 2. \tag{6}$$

We induct on $m$ and divide into the cases of congruence modulo 4 separately. For $m = 1$ equation (2.5) holds. Now suppose the lemma holds for all values less than $4n + 1$ then for $m = 4n + 1$ calculating out by the recurrence relation gives

$$\psi_{4n+1} \equiv (x^2 + A)^{(m^2-1)/4} \mod 2.$$

For $m = 2$ equation (2.6) holds. Now suppose the lemma holds for all values less than $4n + 2$ then for $m = 4n + 2$ calculating out the recurrence relation gives

$$(2y)^{-1}\psi_{4n+2} \equiv (m/2)(x^2 + A)^{(m^2-1)/4} \mod 2.$$

The cases for $m$ congruent to 3 and 4 are completely analogous to these two cases. Thus substituting equations (2.5) and (2.6) into equation (2.4) shows that the right hand side is divisible by 2, and hence $\omega_m$ is in $R$. $\square$

*Remark* 1.6. In the polynomial ring $\mathbb{Z}[A, B, x, y^2]$ we can replace $y^2$ by $x^3 + Ax + B$, and thus take this ring to be $\mathbb{Z}[A, B, x]$.

**Lemma 1.7.** *The highest degree term of $\phi_m(x)$ is $x^{m^2}$ and the highest degree term of $\psi_m^2(x)$ is $m^2 x^{(m^2-1)}$.*

*Proof.* First we prove that

$$\psi_m(x) = \begin{cases} y\left(mx^{(m^2-4)/2} + \cdots\right) & m \text{ even} \\ mx^{(m^2-1)/2} + \cdots & m \text{ odd} \end{cases} \quad (7)$$

by induction. We can immediately see that this holds for $m \leq 4$.

Suppose $m = 2n$, and assume that this formula holds up to $2n$. we look at the case of $n$ even and odd separately. Suppose that equation (2.7) holds for all $m' < 2n$. In the case that $n$ is even, then using the induction hypothesis we get that

$$\psi_{n+2}\psi_n^2 = y\left((n-1)^2(n+1)x^{3n^2/2} + \cdots\right)$$

and

$$\psi_{n-2}\psi_{n+1}^2 = y\left((n-2)(n+1)^2 x^{3n^2/2} \cdots\right)$$

so subtracting gives a leading term of $2x^{3n^2/2}$ (of the univariate in $x$ factor.) Also, the induction hypothesis gives

$$\psi_n = y\left(x^{(n^2-4)/2} + \cdots\right).$$

plugging all this into the recurrence verifies that $\psi_{2n}$ satisfies equation (2.7). In the case that $n$ is odd, then using the induction hypothesis gives that the leading terms are

$$\psi_{n+2}\psi_n^2 = y\left((n-1)^2(n+1)x^{(3n^2-3)/2} + \cdots\right)$$

and

$$\psi_{n-2}\psi_{n+1}^2 = y\left((n-2)(n+1)^2 x^{(3n^2-3)/2} \cdots\right)$$

Also, the induction hypothesis gives

$$\psi_n = nx^{(n^2-1)/2} + \cdots.$$

So combining this all via the recurrence relation gives that $\psi_m$ satisfies the hypothesis.

Now for the case $m = 2n + 1$, and assume that the formula holds up to $2n+1$. In either case, $n$ even or odd, then (after replacing $y^2$ by $x^3 + Ax + B$) the expansions are

$$\psi_{n+1}\psi_n^3 = (n+2)n^3 x^{((2n+1)^2-1)/2} + \cdots$$

and

$$\psi_{n-1}\psi_{n+1}^3 = (n-1)(n+1)^3 x^{((2n+1)^2-1)/2} + \cdots$$

subtracting and plugging into the recurrence relation gives that the expansion is
$$\psi_{2n+1} = (2n+1)x^{((2n+1)^2-1)/2} + \cdots.$$
as desired.

Squaring equation (2.7) (and replacing $y^2$ by $x^3 + Ax + B$) gives the desired equality for $\psi_m^2$.

Now it is a simple matter to show that $\phi_m$ satisfies the statement of the lemma. Using the identity for $\psi_m$ gives that the leading term of $\psi_{m+1}\psi_{m-1}$ is $(m^2-1)x^{m^2}$. Also the leading term of $x\psi_m^2$ is $m^2 x^{m^2}$. Using this in the definition of $\phi_m$ gives that the leading term is $x^{m^2}$ as desired. $\square$

Now that we have thoroughly examined the form of these polynomials, we can give the rational equation for the multiplication by $m$ map.

**Theorem 1.8.** *If $[m]$ denotes the multiplication by $m$ map on $E$, then if the characteristic of $K$ does not divide $m$ then the map is given by rational functions satisfying:*
$$[m](x,y) = \left( \frac{\phi_m(x)}{\psi_m^2(x)}, \frac{\omega_m(x,y)}{\psi_m^3(x,y)} \right).$$

*Proof.* We can take the $x$ coordinate map to be univariate in $x$ by lemma 2.1.7.

We do not present a full proof here in an effort to remain brief. However, there are two ways to prove that this formula is correct. This follows from induction on $m$ and substituting into the addition formula. This approach is very concrete and computational but intricate ([1] III.4.) Alternately, there is an analytic proof that can be found in [23] section 9.5 or [14] section II.1. $\square$

Now that we have defined the multiplication by $m$ map, we can look at the $m$-torsion of $E$, that is the set of points on $E$ with order $m$.

**Definition 1.9.** The *$m$-torsion subgroup* of $E$ is the set of all points in $E(\overline{K})$ with order $m$, and is denoted $E[m]$. Then a point $P$ is in $E[m]$ if and only if $mP$ is the point at infinity.

The following theorem justifies the name of the torsion polynomials.

**Corollary 1.10.** *Suppose that the characteristic of $K$ does not divide $m$. A point $P = (x,y)$ on $E$ is a root of $\psi_m$ if and only if $P$ is an $m$ torsion point.*

*Proof.* By theorem 2.1.8 the $x$ coordinate of $mP$ is given by the map $\frac{\phi_m(x)}{\psi_m^2(x)}$. If $P$ is an $m$ torsion point then the function $\phi_m/\psi_m^2$ has a pole at $P$. This implies that the denominator $\psi_m^2(P)$ is 0, and thus so is $\psi_m(P)$. To show the other direction note that if $\psi_m(P)$ is not 0, then $\phi_m(P)/\psi_m^2(P)$ is a value in $\overline{K}$ corresponding to the $x$-coordinate of $mP$. Thus $mP$ is not the point at infinity, so $P$ is not an $m$ torsion. $\square$

Using this corollary, we can determine the degree of the numerator and denominator of the $x$-coordinate map of $[m]$.

**Lemma 1.11.** *Suppose that the characteristic of $K$ does not divide $m$. Then the polynomials $\phi_m(x)$ and $\psi_m^2(x)$ have no roots in common.*

*Proof.* By definition $\psi_m = x\psi_m^2 - \psi_{m+1}\psi_{m-1}$. Thus if $x$ is a common root of $\phi_m$ and $\psi_m^2$, then x is a root of $\psi_{m+1}\psi_{m-1}$. So reinterpreting this polynomials as functions on $E$ implies that there is some $P$ in $E(\overline{K})$ such that $\phi_m(P)$ and $\psi_m(P)$ are both 0 and $P$ is also a root of $\psi_{m+1}$ or $\psi_{m-1}$. And by corollary 2.1.10 this implies that $P$ is an $m$-torsion, and also an $m+1$ or $m-1$ torsion point. And this implies that either $\gcd(m-1,m)$ or $\gcd(m,m+1)$ is greater than 1. This is a contradiction, thus $\phi_m$ and $\psi_m^2$ must be relatively prime. $\square$

**Lemma 1.12.** *Suppose that the characteristic of $K$ does not divide $m$. Then $E[m]$ is isomorphic to*
$$\mathbb{Z}/m\mathbb{Z} \times \mathbb{Z}/m\mathbb{Z}.$$

*Proof.* By lemma 2.1.7
$$\psi_m^2(x) = m^2 x^{(m^2-1)} + \cdots.$$

So, as the characteristic of $K$ does not divide $m$, $m$ is not zero in $K$ and hence the degree of $\psi_m^2$ is $m^2 - 1$. As it is univariate, this implies that it has $m^2 - 1$ roots. Thus there are $m^2 - 1$ points $P$ in $E(\overline{K})$ such that $\psi_m(P) = 0$. Thus $E[m]$ contains $m^2$ points (including the point at infinity.)

First consider the case that $m$ is prime. Then by the fundamental theorem of finitely generated abelian groups [7] $E[m]$ is isomorphic to either the cyclic group of $m^2$ elements or the direct sum of two copies of $\mathbb{Z}/m\mathbb{Z}$. But by definition, every point in $E[m]$ has order $m$ so $E[m]$ must be $\mathbb{Z}/m\mathbb{Z} \times \mathbb{Z}/m\mathbb{Z}$.

In the case that $m = p^n$, a prime power, this can be seen as follows. The subgroup $E[m]$ must be a direct sum of two cyclic subgroups, if it was not, then the fundamental theorem of finitely generated abelian groups implies that $E[p]$ would be a direct sum of more than two cyclic subgroups, contradicting the result we just showed. Then assume that the theorem holds for $m = p^{n-1}$, then the only two possible isomorphism types for $E[p^n]$ are

$$\mathbb{Z}/p^{n+1}\mathbb{Z} \times \mathbb{Z}/p^{n-1}\mathbb{Z} \text{ or } \mathbb{Z}/m\mathbb{Z} \times \mathbb{Z}/m\mathbb{Z},$$

because, as argued above, $E[m]$ must contain $m^2$ points. However, $E[m]$ must contain only $m$ torsion, and that narrows down the possible isomorphism choices to one.

Thus by the Chinese remainder theorem if $m$ is composite $E[m]$ must also be isomorphic to $\mathbb{Z}/m\mathbb{Z} \times \mathbb{Z}/m\mathbb{Z}$. $\square$

## 2 Isogenies

In the previous section we saw the multiplication by $m$ map, as an example of a map that preserves point addition and the structure of the elliptic curve as a an algebraic variety. Now we will generalize our analysis to account for all such maps, the isogenies.

Various authors treat isogenies in different ways. In [20] Silverman takes an abstract approach favoring an arithmetic geometry point of view. By contrast in [23] Washington takes a more concrete algebraic and computational point of view. In the following presentation, we heavily favor the concrete computational perspective, as the goal is to provide necessary background to understand the algorithms for some of the computational aspects of isogenies. However, to provide deeper insight, when it is not overly distracting we will point out the differences in the more abstract approach to the subject.

We begin with a definition:

**Definition 2.1.** An isogeny $\varphi$ is a nontrivial rational map of an Elliptic Curve onto another Elliptic Curve that is also a group homomorphism.

For those familiar with the the abstract language of category theory, isogenies are the (nontrivial) morphisms in the Category of Elliptic Curves. Indeed, isogenies are rational maps and hence morphisms in the category of algebraic varieties, as well as abelian group homomorphisms. It is worth noting that not all authors even agree on this definition. For example, Silverman allows trivial isogenies, which expands the definition to all morphisms in the category of elliptic curves (defined over a field.) However, for our purposes, we restrict our consideration to the nontrivial case to simplify and avoid having to constantly distinguish the two cases. As an immediate example of how this choice simplifies things we have the following fact:

**Lemma 2.2.** *If $\varphi : E \to E'$ is an isogeny, then $\varphi$ is surjective. Meaning that for a point $P'$ in $E'(\overline{K})$ there exists a point $P$ in $E(\overline{K})$ such that $\varphi(P)$ is $P'$.*

*Proof.* Recall the Theorem of algebraic geometry that all nontrivial mappings of algebraic curves are surjective ([15]: II.6.8). By definition of an isogeny $\varphi$ is a nonzero mapping of algebraic curves, hence it must be surjective. □

Furthermore, Silverman does not define isogenies as group homomorphisms. In that presentation, the definition only requires that $\varphi$ preserves the point at infinity. By looking at the homomorphism that $\varphi$ induces on the principle divisors of $E$, one sees that the property of preserving the point at infinity $\varphi$ implies $\varphi$ is a group homomorphism ([20] Theorem III.4.8.) However, the reader primarily interested in explicit computational methods can easily skip that formalism.

We can take the set $E(\overline{K})$ as an algebraic variety or as a group. When considering $\varphi$ as a map of algebraic varieties, it is denoted $\varphi(x,y)$, and is considered a pair $(x', y')$ satisfying the Weierstrass equation of the codomain. When considering it as a group homomorphism we take $P$ as a general element of $E(\overline{K})$ and denote the evaluation of $\varphi$ on $P$ as $\varphi(P)$, an element of $E'(\overline{K})$ interpreted as a group. In terms of notation we will use $\varphi(P)$ and $\varphi(x,y)$ interchangeably.

With just the basic definitions, we can recognize a couple of examples of isogenies:

*Example* 2.3. Let $m > 0$ be an integer. Suppose the characteristic of $K$ is 0 or relatively prime to $m$, then the the multiplication by $m$ map that sends $P$ to $m \cdot P$ is an isogeny. As we saw above, this map is rational in the coordinates and it maps points $P$ on $E$ to $E$. Furthermore, this multiplication distributes over point addition, and hence is a group homomorphism. Also, $E(\overline{K})$ has infinite order and as argued above the order of the $m$ torsion of $E$ is $m^2$, so multiplication by $m$ cannot annihilate the whole group of points on the curve, and hence is non-constant.

*Example* 2.4. Suppose $K = \mathbb{F}_q$ for a prime power $q = p^n$ then the Frobenius map $(x, y) \mapsto (x^q, y^q)$ is an isomorphism. Clearly from its presentation, this is a rational map in the coordinates. Furthermore, the map $x^q$ distributes over multiplication and addition. Thus if $(x, y)$ satisfies the Weierstrass model: $y^2 + a_1 xy + a_3 y = x^3 + a_2 x^2 + a_4 x + a_6$ with $a_i \in \overline{\mathbb{F}}_p$ then

$$(y^q)^2 + a_1^q(x^q)(y^q) + a_3^q(y^q) = (y^2 + a_1 xy + a_3 y)^q$$
$$= (x^3 + a_2 x^2 + a_4 x + a_6)^q$$
$$= (x^q)^3 + a_2^q(x^q)^2 + a_4^q(x^q) + a_6^q$$

So, given that $\{a_i^q\}_{i=1}^5$ are a-invariants of a non-singular curve $E^q$, the Frobenius mapping: $(x, y) \mapsto (x^q, y^q)$ is a rational map from $E$ to $E^q$.

*Remark* 2.5. Because an isogeny $\varphi$ is a rational map, the evaluation at a point $P$ is given by

$$\varphi(P) = \left(\frac{p_x}{q_x}, \frac{p_y}{q_y}\right)$$

where $p_x, q_x, p_y, q_y$ are polynomials in the coordinates of $P$. Additionally, $p_x$ is relatively prime to $q_x$, $p_y$ is relatively prime to $q_y$, and $p_x$ and $p_y$ are monic. Hence this representation is unique.

**Definition 2.6.** The degree of an isogeny $\varphi$ is the maximum of the degree of the numerator and denominator of the $x$-coordinate maps:

$$\deg(\varphi) = \max\{\deg(p_x), \deg(q_x)\}.$$

## 2.1 Coordinate Maps

At this point, as we are working with rational functions on elliptic curves, it is worthwhile to investigate some basic properties of these maps. To discuss rational functions we must first precisely state what is meant by this. Because an elliptic curve is defined by a Weierstrass equation, the points on the curve satisfy a polynomial equation:

$$W(x, y) = y^2 - x^3 + a_1 xy - a_2 x^2 + a_3 y - a_4 x - a_6 = 0$$

Elliptic curves are irreducible, and hence $W$ is irreducible. Thus the ideal of $\overline{K}[x, y]$ generated by $W$ is prime, so that the quotient ring $R = \overline{K}[x, y]/(W)$ is an integral domain. The rational functions on $E$ are the elements in the field of fractions of $R$, denoted by $\overline{K}(E)$. Occasionally, by abusing notation we will refer to a polynomial on $E$ which means a rational function on $E$ with trivial denominator. Before proving anything about rational functions on $E$, we can first make the following observation about polynomials on elliptic curves:

**Lemma 2.7.** *Let $p(x,y)$ be a polynomial defined on $E$. Then there exists polynomials $p_1(x)$ and $p_2(x)$ both univariate in the $x$-coordinate such that $p(x,y) = p_1(x) + y \cdot p_2(x)$.*

*Proof.* This is shown by induction on the highest degree $m$ of $y$. For the case $m = 1$ we are done. For the case $m = 2$ this can be seen by replacing $y^2$ by according to the Weierstrass equation of $E$. Then assuming that this holds for $n < m$ and substituting via the inductive hypothesis, results in a polynomial where the highest degree power of $y$ is less than $m$. $\square$

We can apply this lemma to simplify the form of rational maps on $E$ that we consider:

**Lemma 2.8.** *Suppose $R(x,y)$ is a rational map on $E$ then there exists polynomials $\phi_1(x), \phi_2(x), \psi(x)$ univariate in $x$ such that*

$$R(x,y) = \frac{\phi_1(x) + y\phi_2(x)}{\psi(x)}.$$

*Proof.* Applying lemma 2.2.7 to $p(x,y)$ and $q(x,y)$ immediately gives that there exists $p_1(x), p_2(x), q_1(x), q_2(x)$ such that:

$$R(x,y) = \frac{p_1(x) + yp_2(x)}{q_1(x) + yq_2(x)}.$$

We can multiply the numerator and denominator through by

$$q_1(x) - (y + a_1 x + a_3) q_2(x).$$

The resulting denominator is

$$(q_1(x))^2 - (y^2 + a_1 x + a_3)(q_2(x))^2 + (a_1 x + a_3) q_1(x) q_2(x)$$
$$= (q_1(x))^2 - (x^3 + a_2 x^2 + a_4 x + a_6)(q_2(x))^2 + (a_1 x + a_3) q_1(x) q_2(x)$$
$$= \psi(x).$$

Applying lemma 2.2.7 to the numerator again gives the desired equality. $\square$

Because $\varphi$ is a group homomorphism, it necessarily preserves negation so $\varphi(-P) = -\varphi(P)$. Recalling the explicit formulas for the coordinates of a negative point:

$$-(x,y) = (x, -y - a_1 x - a_3).$$

Now consider an isogeny $\varphi : E \to E'$, where $E$ and $E'$ are defined by a Weierstrass equations with coefficients $\{a_i\}_{i=1}^{5}$ and $\{a'_i\}_{i=1}^{5}$ respectively. Writing $\varphi(x, y) = (R_1(x, y), R_2(x, y))$ for rational maps $R_1(x, y), R_2(x, y)$ and applying the negation formula it follows that:

$$\varphi(-P) = \varphi(x, -y - a_1 x - a_3)$$
$$= (R_1(x, -y - a_1 x - a_3), R_2(x, -y - a_1 x - a_3)). \quad (8)$$

Likewise:

$$-\varphi(P) = -(R_1(x, y), R_2(x, y))$$
$$= (R_1(x, y), -R_2(x, y) - a'_1 R_1(x, y) - a'_3). \quad (9)$$

These two equations can be combined to greatly simplify the form of the rational maps of isogenies:

**Lemma 2.9.** *If $\varphi$ is an isogeny, then the x-coordinate map of $\varphi$ can be expressed as a univariate (in x) rational map: $r_1(x)$.*

*Proof.* First we apply lemma 2.2.8 to $R_1(x, y)$ so that we have univariate (in $x$) polynomials $\phi_1, \phi_2, \psi$ with

$$R_1(x, y) = \frac{\phi_1(x) + y\phi_2(x)}{\psi(x)}.$$

Combining this with equations (2.8) and (2.9) gives:

$$\frac{\phi_1(x) + y\phi_2(x)}{\psi(x)} = \frac{\phi_1(x) - (y + a_1 x + a_3)\phi_2(x)}{\psi(x)}.$$

Then subtracting the right hand side from the left hand side it follows that:

$$\frac{(2y + a_1 x + a_3)\phi_2(x)}{\psi(x)} = 0.$$

The polynomial $2y + a_1 x + a_3$ is the two torsion polynomial for $E$ thus only satisfied at two torsion points $P$. Therefore for this polynomial to be satisfied at all points $P = (x, y)$ we must necessarily have $\phi_2(x) = 0$. Thus

$$R_1(x, y) = \frac{\phi_1(x)}{\psi(x)}.$$

□

**Lemma 2.10.** *If $\text{char}(K) \neq 2$, then the y-coordinate map of $\varphi$ is of the form:*

$$(y + (a_1 x + a_3)/2)\, r_2(x) - (a'_1 r_1(x) + a'_3)/2$$

*where $r_2(x)$ is a univariate rational map and the x-coordinate map of $\varphi$ is given by $r_1(x)$.*

*Proof.* Similar to the proof of lemma 2.2.9, first apply lemma 2.2.8 to $R_1(x, y)$ so that we have univariate (in $x$) polynomials $\phi_1, \phi_2, \psi$ with

$$R_2(x, y) = \frac{\phi_1(x) + y\phi_2(x)}{\psi(x)}.$$

From lemma 2.2.9 we have that the $x$-coordinate map of $\varphi$ is a univariate rational map $r_1(x)$. Combining both of these facts with equations (2.8) and (2.9) gives the equality:

$$\frac{\phi_1(x) - (y + a_1 x + a_3)\phi_2(x)}{\psi(x)} = -\frac{\phi_1(x) + y\phi_2(x)}{\psi(x)} - (a'_1 r_1(x) + a'_3). \quad (10)$$

Straight forward algebraic manipulation (because we are not in characteristic 2) gives:

$$\frac{\phi_1(x)}{\psi(x)} = \frac{(a_1 x + a_3)\phi_2(x)}{2\psi(x)} - (a'_1 r_1(x) + a'_3)/2.$$

Then we can substitute this into equation (2.10), and solve for $R_2(x, y)$ which gives:

$$R_2(x, y) = (y + (a_1 x + a_3)/2)\frac{\phi_2(x)}{\psi(x)} - (a'_1 r_1(x) + a'_3)/2$$
$$= (y + (a_1 x + a_3)/2)\, r_2(x) - (a'_1 r_1(x) + a'_3)/2,$$

where $r_2(x)$ is a univariate rational map (in $x$), as desired. □

## 2.2 Separability

With the results we've proved about the explicit forms of the rational maps that occur as coordinate maps in isogenies, we can now discuss some further properties of isogenies.

**Definition 2.11.** Let $\varphi : E \to E'$ be an isogeny, and let $r_1(x)$ be the $x$-coordinate map. If the derivative of the $x$-coordinate map $r'_1(x)$ is not 0 then $\varphi$ is *separable*.

With this definition it is instructive to look at some examples:

*Example* 2.12. Suppose $\mathbb{F}_p$ is a finite field with prime order $p$, and let $E/\mathbb{F}_p$ be an elliptic curve. Then the Frobenius isogeny is given by the rational maps $(x^p, y^p)$. So the derivative of the $x$-coordinate map is,

$$p \cdot x^{p-1} = 0$$

because this in characteristic $p$. Thus by definition the Frobenius isogeny is not separable.

*Example* 2.13. Suppose $E/\mathbb{Q}$, and $\varphi : E \to E_2$ is an isogeny, and furthermore inseparable. Then if $r(x)$ is the $x$-coordinate map of $\varphi$ then $r'(x) = 0$ so $r(x)$ is constant. Which cannot be, hence $\varphi$ cannot be inseparable.

These two examples show two extremes. In the case of elliptic curves over $\mathbb{Q}$ isogenies are always separable. On the other hand, in the finite field case Frobenius isogenies are always not separable.

It is instructive to note that there is an alternate (yet equivalent) approach to the defining the separable property. Silverman prefers a characterization of separability based on function field extensions. Specifically, an isogeny is a non-constant map of algebraic curves, so it induces an injection between the corresponding function fields

$$\varphi^* : \overline{K}(E_2) \to \overline{K}(E_1)$$

by precomposing functions in $\overline{K}(E_2)$ with the isogeny $\varphi$. Then $\overline{K}(E_1)$ is an extension of the field $\varphi^*\overline{K}(E_2)$ see ([20] Theorem II.2.4). Using this language and notation, an isogeny is separable if the corresponding extension of fields $\overline{K}(E_1)/\varphi^*\overline{K}(E_2)$ is separable. Although the more computationally minded reader may find this overly abstract, if one is comfortable with algebraic geometry this equivalent characterization is useful to keep in mind.

Using this definition of separability, it is also instructive to look at the interplay between $\varphi$ as a rational map and as a group homomorphism. Because an isogeny is a rational map, if either of the denominators of the coordinate maps evaluates to 0, the result of the isogeny will be the point at infinity. Intuitively, one can think of this as dividing either of the coordinates by 0 will send the point to infinity. More formally, dividing a coordinate by 0 indicates that the corresponding point in the projective plane is $(0 : 1 : 0)$, the point at infinity. Hence, the kernel corresponds to points that form the roots of the denominator polynomials. By lemma 2.2.8 the denominator polynomials are univariate in the $x$-coordinate, and hence have a finite number of roots. Thus the kernel of an isogeny is a finite subgroup of $E(\overline{K})$, with order bounded by the degree of the isogeny.

Thus we can classify isogenies based on the relation of the order of the kernels and the degrees as rational maps:

**Lemma 2.14.** *If $\varphi : E \to E'$ is a separable isogeny then $|\ker(\varphi)| = \deg(\varphi)$. Otherwise $|\ker(\varphi)| < \deg(\varphi)$.*

*Proof.* By lemma 2.2.2, we know that $\varphi$ is surjective. So if $P = (a, b) \in E'(\overline{K})$ and $P$ not $\infty$ then there exists $(x_0, y_0) \in E(\overline{K})$ such that $(a, b) = \varphi(x_0, y_0)$. By lemma 2.2.9 we have that

$$r_1(x_0) = \frac{p(x_0)}{q(x_0)} = a.$$

Furthermore, because $E'(\overline{K})$ is infinite we can choose $(a, b)$ with the following properties:

1. $a \neq 0$

2. $\deg(p(x) - aq(x)) = \max\{\deg(p(x)), \deg(q(x))\} = \deg(\varphi)$ (The only way that $\deg(p(x) - aq(x)) < \deg(\varphi)$ is possible is if $\deg(p(x)) = \deg(q(x))$ and $\alpha$ is the leading coefficient of $p$, $\beta$ is the leading coefficient of $q$, and $\alpha - a\beta = 0$. In this case, we only need restrict $a \neq \alpha/\beta$.

We have that $\deg(p(x) - aq(x)) = \deg(\varphi)$, and hence has $\deg(\varphi)$ (possibly indistinct) roots. As $\varphi$ is a homomorphism the number of distinct roots of $p(x) - aq(x)$ is exactly $|\ker(\varphi)|$. Now it suffices to determine when $p(x) - aq(x)$ has repeated roots. The polynomial $p(x) - aq(x)$ has repeated roots at $x_0$ if and only if $p(x_0) - aq(x_0) = 0$ and $p'(x_0) - aq'(x_0) = 0$. In this case, $ap'(x_0)q(x_0) = ap(x_0)q'(x_0)$. We chose $a \neq 0$ so this implies that $x_0$ is a root of $p(x)'q(x) - p(x)q'(x)$. Furthermore, $r_1'(x) = 0$ and hence $r_1$ is by definition inseparable if and only if $p(x)'q(x) - p(x)q'(x) = 0$ for all $x \in \overline{K}$.

So if $\varphi$ is not separable, then every element of $\overline{K}$ is a root of $p(x)'q(x) - p(x)q'(x)$ and hence $p(x) - aq(x)$ must have a repeated root.

If $\varphi$ is separable, then $p'(x)q(x) - p(x)q'(x)$ is not identically 0, and hence has a finite number of roots. We can let $S$ be this finite set of roots and further restrict our choice of $a$ so that $a \notin r_1(S)$. As such, if $x_0$ were a repeated root of $p(x) - aq(x)$ then the preceding argument shows that $x_0 \in S$, a contradiction. Thus, in the separable case, we conclude that $\deg(\varphi) = |\ker(\varphi)|$. □

## 2.3 Isogenies and Differential Forms

An important tool in the study of elliptic curves are the differentials of the function field $\overline{K}(E)$. Similarly, this tool is also important to the study of isogenies. Recall the definition ([20] II.4)

**Definition 2.15.** The *space of differential forms* of $E$, denoted $\Omega_E$ is the 1-dimensional $\overline{K}(E)$-vector space generated by $dx$. Here $df$ is the usual differential operator, such that given $f$ and $g$ in $\overline{K}(E)$ and $a$ is constant in $\overline{K}$

1. $d(f + g) = df + dg$.
2. $d(fg) = f\,dg + g\,df$.
3. $da = 0$.

Using this space, applying the differential operator to the Weierstrass equation for $E$ gives us the following important value associated to $E$ ([20] III.5):

**Definition 2.16.** The *invariant differential* of $E$, denoted $\omega$ is the value:

$$\omega = \frac{dx}{2y + a_1x + a_3} = \frac{dy}{3x^2 + 2a_2x + a_4 - a_1y}.$$

We want to understand the effect of $f$ under mappings of $E$ so to this end we make the following definition

**Definition 2.17.** Let $f$ be a map from $E$ to a curve $E'$ where $f_x$ and $f_y$ are the $x$ and $y$ coordinate maps, respectively. Let $\gamma$ be a differential form on $E'$, hence $\gamma = \alpha dx'$, for some $\alpha$ in $\overline{K}(E')$ Then the *pullback* of $\gamma$ along $f$ is denoted $f^*\gamma$ and is defined as
$$\alpha(f_x, f_y)df_x.$$

*Remark* 2.18. The map $f_x$ is a function on $E$, so $df_x$ is in $\Omega_E$. Thus, $f^*$ defines a mapping from $\Omega_{E'}$ to $\Omega_E$.

Ultimately we want to use differential forms to study isogenies. However, it is prudent to look at the effect of another type of map on the invariant differential. Let $Q$ be any point on a curve $E$ and define $t_Q : E \to E$ as the translation by $Q$ map, specifically, $t_Q(P) = P + Q$. It is useful to the study of isogenies to understand the pullback of the invariant differential $\omega$ along $t_Q$. It turns out that the invariant differential, is in fact, invariant under translation (hence the name.)

**Lemma 2.19.** *For any point $Q$ on $E$ the pullback of the invariant differential $\omega$ along the translation map $t_Q$ is $\omega$.*

*Proof.* This can be seen by writing out the addition formulas on $E$ and a straight forward algebraic manipulation confirms that
$$\frac{dx_{P+Q}}{2y_{P+Q} + a_1 x_{P+Q} + a_3} = \frac{dx_P}{2y_P + a_1 x_P + a_3},$$
for any $P$ and $Q$ on $E$.

There is a more elegant alternate proof ([20] III.5.1) that uses the effect of $t_Q^*$ on the divisor of $\omega$. That proof may be more elucidating for readers familiar with algebraic geometry. □

We can begin to discuss the pullback of invariant differentials along isogenies. First we can immediately see that the invariant differential of $E'$ pulls back to the invariant differential of $E$.

By applying this fact, we can precisely determine the pullback of the invariant differential along an isogeny of $E$.

**Lemma 2.20.** *Suppose $K$ is of characteristic not equal to 2. If $\varphi : E \to E'$ is an isogeny and $\omega'$ is the invariant differential of $E'$, then $\varphi^*\omega' = c\omega$ for some constant $c$ in $\overline{K}$.*

*Proof.* By considering $\varphi^*\omega'$ as an element in $\Omega_E$ there is a $g$ in $\overline{K}(E)$ such that $\varphi^*\omega = g\omega$. Also $t_Q^*\varphi^*\omega' = \varphi^* t_{\varphi(Q)}^*\omega'$, because $\varphi$ is a group homomorphism.

By lemma 2.2.19 it follows that $t_{\varphi(Q)}^*\omega' = \omega'$ and $t_Q^*\omega = \omega$. Hence for all $Q$ on $E$ we have:
$$t_Q^* g = \frac{\varphi^* t_{\varphi(Q)}^*\omega'}{t_Q^*\omega'} = \frac{\varphi^*\omega'}{\omega}.$$

Thus $g$ must be constant, so $\varphi^*\omega = c\omega$ for some $c$ in $\overline{K}$. □

Recall that in lemmas 2.2.9 and 2.2.10 we greatly simplified the form of the $x$ and $y$ coordinate maps of $\varphi$. Specifically, we showed that the $\varphi_x$ map is univariate in the $x$-coordinate of $E$, and also we expressed the map $\varphi_y$ in terms of the $y$ coordinate and Weierstrass coefficients of $E$ as well as $\varphi_x$ and some other rational function, univariate in $x$. Using the identity for $\varphi^*\omega$, we can entirely express this other univariate rational function in the map $\varphi_y$ in terms of $\varphi_x$.

**Lemma 2.21.** *Suppose $\varphi : E \to E'$ is an isogeny with coordinate maps $\varphi_x(x)$ and $\varphi_y(x, y)$ then*

$$\varphi_y(x, y) = c \left( y + \frac{a_1 x + a3}{2} \right) \varphi'_x(x) - \frac{a_1 \varphi_x(x) + a_3}{2},$$

*where $c$ is some constant in $\overline{K}$ and $\varphi'_x(x)$ denotes the derivative of $\varphi_x$ with respect to $x$, as usual.*

*Proof.* By lemma 2.2.20, we have that

$$\frac{d\varphi_x}{2\varphi_y + a_1 \varphi_x + a_3} = \frac{c\,dx}{2y + a_1 x + a_3}.$$

Next we note that $d\varphi_x = \varphi'_x dx$, substituting this in and solving for $\varphi_y(x, y)$ gives the desired equality. □

Now that we have determined the general form of the coordinate maps of an isogeny we can characterize the isogenies based on the constant multiple in the $y$-coordinate map.

**Definition 2.22.** An isogeny $\varphi : E \to E'$ is *normalized* if the pullback of the invariant differential of $E'$ along $\varphi$ is equal to the invariant differential of $E$. That is, if $\omega$ and $\omega'$ are the invariant differentials of $E$ and $E'$ respectively, then $\varphi^*\omega'$ equals $\omega$.

## 2.4 The Dual Isogeny

In the final section on the basic theory of elliptic curve isogenies, we examine the question: Suppose there is a degree $\ell$ isogeny from $E_1$ to $E_2$, is there a degree $\ell$ isogeny from $E_2$ back to $E_1$? The answer is yes. Not only does such a map exist, but there exists a unique such map satisfying some nice properties. Here we only prove and state the result for separable isogenies, but it does in fact hold for all isogenies.

**Theorem 2.23.** *Let $\varphi : E_1 \to E_2$ be a separable isogeny of degree $\ell$. Then there exists a unique separable isogeny $\hat{\varphi} : E_2 \to E_1$ of degree $\ell$ such that $\hat{\varphi} \circ \varphi$ is the multiplication by $\ell$ map on $E_1$. The isogeny $\hat{\varphi}$ is called the dual of $\varphi$.*

*Proof.* In an effort to remain brief, we present only a high level proof here and do not delve into the background details.

We suppose that $\varphi$ is separable so the characteristic of $K$ does not divide $\ell$. Thus $|\ker(\varphi)|$ is $\ell$ by lemma 2.2.14.

By [20] corollary III.4.11, if $\phi : E_1 \to E_2$ and $\psi : E_1 \to E_3$ are isogenies and $\phi$ is separable with $\ker(\psi)$ containing $\ker(\phi)$ then there is a unique isogeny $\lambda : E_2 \to E_3$ such that $\psi = \lambda \circ \phi$. (The proof of this works by using [20] theorem III.4.10(c) to generate a tower of Galois extensions: $\psi^*\overline{K}(E_3) \subseteq \psi^*\overline{K}(E_2) \subseteq \overline{K}(E_1)$ and deducing the existence of $\lambda$ from this.)

Thus in this case, this result gives that there exists a unique isogeny $\hat{\varphi} : E_2 \to E_1$ such that $\hat{\varphi} \circ \varphi = [\ell]$. Furthermore, we know that $|E[\ell]|$ is $\ell^2$ by lemma 2.1.12. As isogenies are group homomorphisms, $\varphi(E[\ell]) \cong E[\ell]/\ker(\varphi)$. Hence $|\varphi(E[\ell])|$ is $\ell$. Also $\varphi(E[\ell]) = \ker(\hat{\varphi})$ as $E[\ell]$ is exactly the kernel of multiplication by $\ell$. So thus the order of $\ker(\hat{\varphi})$ is $\ell$.

It follows that $\hat{\varphi}$ must be separable. As $[\ell]$ is $\hat{\varphi} \circ \varphi$ and $[\ell]$ has degree $\ell^2$ and $\varphi$ has degree $\ell$, then $\hat{\varphi}$ has degree $\ell$. Thus by lemma 2.2.14 is separable.

For a complete proof, including the case of inseparable isogenies see [20] theorem III.6.1. There is a complete proof in [23], theorem 12.14, that is more computational. The proof presented above is a hybrid of the two approaches. □

*Remark* 2.24. From the proof of theorem 2.2.23 it follows that $\varphi \circ \hat{\varphi}$ is the multiplication by $\ell$ map on $E_2$. Furthermore, $\hat{\hat{\varphi}} = \varphi$.

# Part III
# Algorithms

Before examining the algorithms for computing isogenies, it is prudent to examine what exactly this means. In one sense, as an isogeny is a function, computing it means to evaluate it on some input. By definition isogenies are rational maps, so given this rational map it is straight forward to perform this evaluation. However, the rational maps are not the only way to represent an isogeny. In this chapter we give two methods for computing the rational maps of an isogeny. First we assume that we know the domain and kernel, and give algorithms for determining the codomain and rational maps. We also assume that we know the domain, codomain and degree of an isogeny, and we show how to recover the kernel and hence rational maps.

## 3 Computing from the Kernel

Suppose that one knows the kernel and the domain of a separable normalized isogeny. The algorithms in this section show how to compute the codomain and the rational maps associated to that isogeny. Even this computational task is complicated by the ambiguity of representing the kernel. There are two choices. First we can consider the kernel as a list of points in $E(\overline{K})$. Alternately, we can assume that the kernel $C$ is specified by the *kernel polynomial*, the unique monic polynomial of lowest degree with roots only at $x$-coordinates of the finite points of $C$. Vélu's formulas take as input the kernel as a list of points, and return the rational maps and codomain of the curve. Kohel's approach takes the input as the kernel polynomial.

Before going into the details of the algorithms that compute an isogeny given the kernel, it is useful to look at just exactly what can be computed from this input. Specifically, we have to consider post composition of an isogeny by curve isomorphisms (and automorphisms.)

Suppose that $\varphi : E \to E'$ is a separable isogeny with kernel $C$. Also, suppose that $\rho : E' \to E''$ is an isomorphism of curves defined over $K$ (a separable isogeny of degree 1.) Then, $\rho \circ \varphi$ is a separable isogeny from $E$ to $E''$. Thus it is clear that the codomain of an isogeny is not uniquely specified by the kernel.

Now suppose that $\rho : E \to E'$ is a normalized isomorphism of curves (again here isomorphism denotes a separable isogeny of degree 1.) So the invariant differential of $E'$ pulls back to the invariant differential of $E$. Because $\rho$ is a separable degree 1 isogeny it is a linear change of variables, this implies that

$$\rho(x,y) = (x + r, y + sx + (sr + t))$$

for some $r$, $s$ and $t$ in $\overline{K}$. Hence it follows that the $c$-invariants of $E$ and $E'$ are the same, so $E$ and $E'$ are the same curve. Thus, if $E$ and $E'$ are isomorphic but not equal elliptic curves and $\tau : E \to E'$ is an isomorphism of elliptic curves, then $\tau$ is not normalized. So post composing a separable normalized isogeny by an isomorphism to a different elliptic curve results in a non normalized isogeny. It follows that the kernel uniquely specifies the codomain of a separable normalized isogeny.

It remains to consider post composition of an isogeny by automorphisms. That is degree 1 isogenies from a curve $E$ to itself. If the characteristic of $K$ is not 2 or 3 then by ([20] theorem III.10.1 and proposition A.1.2) any automorphism of $E$ is of the form:

$$(x,y) \mapsto (u^2 x, u^3 y)$$

for some $u$ in $\overline{K}$. Hence, any nontrivial automorphism will have $u$ not 1, and then by ([20] section III.1) the pullback of the invariant differential along a nontrivial automorphism will introduce a factor of $u$. Thus post composing a separable normalized isogeny by a nontrivial automorphism will result in a non-normalized isogeny. (This is also the case if the characteristic of $K$ is 3 and the $j$-invariant of $E$ is not 0.) In the case of characteristic $K$ equal to 2 or 3 then there are additional concerns because the automorphisms do not always have such a simple form. Namely, there are nontrivial automorphisms of $E$ under which the pullback of the invariant differential does introduce a scaling factor (see the proof of proposition A.1.2 in [20] for the specific cases.)

This shows that there are cases where post composition of a separable normalized isogeny with an automorphism, results in another separable and normalized isogeny with the same kernel and codomain. This indicates that the kernel and codomain cannot uniquely specify the rational maps for evaluating a seperable normalized isogeny.

Bearing this in mind, as in [4] we remark that given a kernel of a separable normalized isogeny we can uniquely determine a codomain curve (and give a Weierstrass equation for it.) However, we can only specify the rational maps for evaluating the isogeny up to post composition with an automorphism.

## 3.1 Vélu's Approach: Computing from points in the kernel

Vélu's formulas show how, for any field $K$, given a Curve $E_1/K$ and the Kernel of an isogeny (as a list of the points of a finite order subgroup of $E(\overline{K})$) how to determine the codomain of the isogeny, as well as compute the isogeny.

**Input:** Given a curve in general Weierstrass form:

$$E_1 : y^2 + a_1 xy + a_3 y = x^3 + a_2 x^2 + a_4 x + a_6,$$

and a set of points of $C$ that forms a finite subgroup of $E_1(\overline{K})$.

**Output:** The general Weierstrass coefficients of a Weierstrass model for the codomain curve $E_2$ of a separable normalized isogeny with kernel $C$. Also, coordinate maps (as rational maps on $E_1$) that evaluate a point $(x, y)$ on $E_1$ to a point on $E_2$.

**Step 1:** Partition the set of points $C$ :

1. Throw out $\infty$.

2. Let $C_2$ be all the 2-torsion points in $C$, let $R$ be the rest of the points in $C$.

3. Split $R$ into two equal sized sets such that $R_+$ and $R_-$ so that if a point $P$ is in $R_+$ then $-P$ is in $R_-$.

4. Let $S = R_+ \cup C_2$.

**Step 2:** Now given $Q \in S$ define the following quantities:

$$g_Q^x = 3x_Q^2 + 2a_2 x_Q + a_4 - a_1 y_Q$$
$$g_Q^y = -2y_Q - a_1 x_Q - a_3$$
$$v_Q = \begin{cases} g_Q^x & \text{if } 2Q = \infty \\ 2g_Q^x - a_1 g_Q^y & \text{otherwise} \end{cases}$$
$$u_Q = (g_Q^y)^2$$
$$v = \sum_{Q \in S} v_Q, \quad w = \sum_{Q \in S} (u_Q + x_Q v_Q)$$

**Step 3:** Compute the target image:

First define the values:
$$A_1 = a_1, \ A_2 = a_2, \ A_3 = a_3,$$
$$A_4 = a_4 - 5v, \ A_6 = a_6 - (a_1^2 + 4a_2)v - 7w.$$

Then the Weierstrass equation of $E_2$ is:
$$y^2 + A_1 xy + A_3 y = x^3 + A_2 x^2 + A_4 x + A_6.$$

**Step 4:** The formula for computing the image point $(\alpha, \beta)$ from the point $(x, y)$:

$$\alpha = x + \sum_{Q \in S} \left( \frac{v_Q}{x - x_Q} - \frac{u_Q}{(x - x_Q)^2} \right) \tag{11}$$

$$\beta = y - \sum_{Q \in S} \left( u_Q \frac{2y + a_1 x + a_3}{(x - x_Q)^3} + v_Q \frac{a_1(x - x_Q) + y - y_Q}{(x - x_Q)^2} + \frac{a_1 u_Q - g_Q^x g_Q^y}{(x - x_Q)^2} \right) \tag{12}$$

*Remark* 3.1. Note that while Vélu's formulas clearly can be used to evaluate an isogeny (given the domain and kernel) at a given point of the domain curve, here we are treating Vélu's formulas as a way to precompute the rational maps of the isogeny. These rational maps can be stored and used to evaluate any number of points on the domain curve.

For a full proof of the correctness of these algorithms one can see Vélu's original paper [22], or the reader more familiar with English can read Washington's treatment [23] which deals only with the case of characteristic not equal to 2. Here we will partially prove that these formulas work, stating one lemma without proof:

**Lemma 3.2.** *The codomain curve $E_2$ found by Vélu's formulas is nonsingular.*

The proof requires the Riemann-Hurwitz theorem ([15] section IV.2) This is a result from algebraic geometry that relates the degree of an unramified rational map on algebraic curves with the genus of the domain and codomain curves. The proof of the Riemann-Hurwitz theorem requires considerably more depth of understanding of algebraic geometry (particularly it uses sheaves) than most of the other results presented herein. As such, these details are left up to readers who are more interested in these abstract matters.

Furthermore, as is common with many algorithmic results, those interested solely in computation can skip this result, or take it without proof. Indeed, every time one runs Vélu's formulas, one can perform a check that the calculated codomain is in fact nonsingular. Thus in effect proving that the codomain is nonsingular every time the algorithm is run.

The rest of the proof closely follows the presentation in [23]. Before proving the the main theorem we first prove the following lemma:

**Lemma 3.3.** *The rational maps $\alpha(P)$ and $\beta(P)$ in Vélu's formulas obey the following formulas*

$$\alpha(P) = x_P + \sum_{Q \in C - \{\infty\}} (x_{P+Q} - x_Q)$$

and

$$\beta(P) = y_P + \sum_{Q \in C - \{\infty\}} (y_{P+Q} - y_Q)$$

*Proof.* We will make heavy use of the addition formula in section [20] algorithm III.2.3 and restated here in section 2.1.

First we will assume that $Q \in C - \{\infty\}$ is a two torsion point and $P$ us any point on $E_1$ distinct from $Q$. Furthermore, because $Q$ is a two torsion point $g_Q^y = 0$ (as $g_Q^y$ is the bivariate two torsion polynomial evaluated on the point $Q$). This also implies that $u_Q$ is also 0. Substituting the addition formula and expanding in terms of $x_P$ and $y_P$ gives:

$$\begin{aligned}x_{P+Q} =& (y_P^2 - 2y_P y_Q + y_Q^2 + a_1 x_P y_P \\ & - a_1 x_P y_Q - a_1 x_Q y_P + a_1 x_Q y_Q - a_2 x_P^2 \\ & + 2a_2 x_P x_Q - a_2 x_Q^2 - x_P^3 + x_P x_Q^2 \\ & + x_P^2 x_Q - x_Q^3)/(x_P - x_Q)^2.\end{aligned}$$

We can substitute:

$$y^2 - x^3 + a_1 xy = -a_3 y + a_4 x + a_6.$$

for $(x, y)$ any point on $E_1$ and also

$$-2y_P y_Q - a_1 x_Q y_P - a_3 y_P = y_P(g_Q^y) = 0.$$

Then by subtracting $x_Q$ from the resulting expression for $x_{P+Q} - x_Q$, the denominator becomes:

$$a_4 x_P - a_3 y_Q + a_4 x_Q + 2a_6 = -a_1 x_P y_Q + 2a_2 x_P x_Q + 3x_P x_Q^2 + x_Q^3.$$

Using the Weierstrass equation to make a substitution for the terms uniform in the coordinates of $Q$ the whole expression becomes

$$x_{P+Q} - x_Q = \frac{(x_P - x_Q)v_Q + y_Q g_Q^y}{(x_P - x_Q)^2}.$$

Thus this simplifies down to:

$$x_{P+Q} - x_Q = \frac{v_Q}{x_P - x_Q} \tag{13}$$

Now we do a similar evaluation in terms of the $y$ coordinates when translating by $Q$. Substituting the addition formula for the $y$-coordinate, and using the equality

$$x_{P+Q} = \frac{v_Q}{x_P - x_Q} + x_Q$$

gives

$$y_{P+Q} - y_Q = -\frac{y_P - y_Q + a_1(x_P - x_Q)}{x_P - x_Q}\left(\frac{v_Q}{x_P - x_Q} + x_Q\right)$$
$$+ \frac{-y_P x_Q + y_Q x_P - a_3 x_P + a_3 x_Q - y_Q x_P + y_Q x_Q}{x_P - x_Q}$$
$$= -v_Q \frac{a_1(x_P - x_Q) + y_P - y_Q}{(x_P - x_Q)^2} - g_Q^y.$$

This simplifies to:

$$y_{P+Q} - y_Q = -v_Q \frac{a_1(x_P - x_Q) + y_P - y_Q}{(x_P - x_Q)^2}. \tag{14}$$

Next we note that if $P = Q$ then $x_{P+Q} = y_{P+Q} = \infty$ as $Q$ is a two torsion. Then $x_P = x_Q$, so $x_P - x_Q = 0$ so $v_Q/(x_P - x_Q) = \infty$ and hence in equations (3.3) and (3.4) both sides of the equations go off to infinity. Thus these equations hold when $P = Q$ as well.

Now we prove similar results for the case that $Q$ is not a two torsion, and $P$ is not $\pm Q$. In this case, we need to keep track of multiple different addition formulas, so we denote $\lambda_{P+Q}$ and $\nu_{P+Q}$ as the values in the addition formula for $P + Q$. Similarly we define $\lambda_{P-Q}$ and $\nu_{P-Q}$ while computing $P - Q$. Furthermore, these values are related in the following way:

$$\lambda_{P-Q} = \lambda_{P+Q} - g_Q^y, \quad \nu_{P-Q} = \nu_{P+Q} + \frac{x_P g_Q^y}{x_P - x_Q}.$$

In the case of the $x$-coordinates we have that $x_Q = x_{-Q}$ so that:

$$x_{P+Q} - x_Q + x_{P-Q} - x_{-Q} = x_{P+Q} + x_{P-Q} - 2x_Q.$$

Then expanding in terms of $x_P$, $x_Q$, $\lambda_{P+Q}$ and $g_Q^y$ gives:

$$x_{P+Q} + x_{P-Q} - 2x_Q = 2\lambda_{P+Q} + 2a_1\lambda_{P+Q} - 2a_2 - 2x_P - 4x_Q$$
$$- \frac{2g_Q^y \lambda_{P+Q}}{x_P - x_Q} - \frac{a_1 g_Q^y}{x_P - x_Q} + \frac{v_Q}{x_P - x_Q}.$$

Then further expanding in terms of $x_P$ and $x_Q$ gives that

$$2\lambda_{P+Q} + 2a_1\lambda_{P+Q} - 2a_2 - 2x_P - 4x_Q - \frac{2g_Q^y \lambda_{P+Q}}{x_P - x_Q} - \frac{a_1 g_Q^y}{x_P - x_Q}$$
$$= (a_1^2 x_P x_Q + 6x_P x_Q^2 + 4a_2 x_P + a_1 a_3 x_P + 2a_4 x_P$$
$$- a_1^2 x_Q^2 - 6x_Q^3 - 4a_2 x_Q^2 - a_1 a_3 x_Q - 2a_4 x_Q)/(x_P - x_Q)^3.$$

Then replacing $2a_4 x_P$ and $2a_4 x_Q$ via the Weierstrass equation this whole expression simplifies down to $u_Q/(x_P - x_Q)^2$. Combining this all gives that:

$$x_{P+Q} - x_Q + x_{P-Q} - x_{-Q} = \frac{v_Q}{x_P - x_Q} + \frac{u_Q}{(x_P - x_Q)^2}. \tag{15}$$

Next we compute a similar equality for the $y$-coordinates. First we note that the inversion formula gives:

$$y_{-Q} - y_Q - a_1 x_Q - a_3.$$

Then substituting this gives

$$y_{P+Q} - y_Q = y_{P-Q} - y_{-Q} = y_{P+Q} - y_{P-Q} + a_1 x_Q + a_3.$$

Furthermore, $g_P^y = -2y_P - a_1 x_P - a_3$ by slightly abusing notation, because $P$ is not necessarily in $C$. So using this equality and the addition formulas, this expression becomes

$$-(\lambda_{P+Q} + a_1)\frac{v_Q}{x_P - x_Q} - (\lambda_{P+Q} + a_1)\left(\frac{u_Q}{(x_P - x_Q)^2} + 2x_Q\right)$$
$$+ \frac{g_Q^y}{x_P - x_Q} x_{P+Q} + u_Q \frac{g_P^y}{(x_P - x_Q)^3}$$
$$- 2\nu_{P+Q} - \frac{x_P g_Q^y}{(x_P - x_Q)} + a_1 x_Q - a_3.$$

Now by expressing $\lambda_{P+Q}$, $\nu_{P+Q}$, $x_{P+Q}$, $g_Q^x$, and $g_Q^y$ as expressions in $x_P$, $y_P$, $x_Q$ and $y_Q$ one can see that

$$-\frac{a_1 u_Q - g_Q^x g_Q^y}{(x_P - x_Q)^2} = -(\lambda_{P+Q} + a_1)\left(\frac{u_Q}{(x_P - x_Q)^2} + 2x_Q\right)$$
$$- 2\nu_{P+Q} - \frac{x_P g_Q^y}{(x_P - x_Q)} + a_1 x_Q - a_3$$
$$+ g_Q^y (f(Q) - f(P)).$$

The function $f$ in $K(E_1)$ is
$$f(x,y) = y^2 - x^3 + a_1 xy - a_2 x^2 + a_3 y - a_4.$$

The Weierstrass equation gives that for any point $P$ on $E_1$ the evaluation $f(P) = -a_6$. So $f(Q) - f(P)$ is 0. Substituting this all back together gives

$$y_{P+Q} - y_Q + y_{P-Q} - y_{-Q} = -u_Q \frac{2y_P + a_1 x_P + a_3}{(x_P - x_Q)^3} - v_Q \frac{a_1(x_P - x_Q) + y_P - y_Q}{x_P - x_Q}$$
$$- \frac{a_1 u_Q - g_Q^x g_Q^y}{(x_P - x_Q)^2}. \tag{16}$$

When $Q$ is a two torsion point, then we argued that $g_Q^y = u_Q = 0$, thus by partitioning $S$ into the disjoint sets $C_2$ and $R_+$ and substituting the results of equations (3.3) and (3.5), we can evaluate the see that the sum in the $\alpha$ map:

$$\sum_{Q \in S} \frac{v_Q}{x_P - x_Q} - \frac{u_Q}{(x_P - x_Q)^2} = \sum_{Q \in C - \{\infty\}} x_{P+Q} - x_Q.$$

Likewise, by using the results of equations (3.4) and (3.6) we can evaluate the sum in the $\beta$ map:

$$-\sum_{Q \in S} u_Q \frac{2y_P + a_1 x_P + a_3}{(x_P - x_Q)^3} + v_Q \frac{a_1(x_P - x_Q) + y_P - y_Q}{x_P - x_Q} + \frac{a_1 u_Q - g_Q^x g_Q^y}{(x_P - x_Q)^2}$$
$$= \sum_{Q \in C - \{\infty\}} y_{P+Q} - y_Q.$$

This gives the statement of the lemma. □

Now using this proof we finally prove the following theorem.

**Theorem 3.4.** *Steps 1-4 of Vélu's formulas give the domain and rational maps to compute a separable normalized isogeny with kernel $C$.*

*Proof.* Define $t = x/y$ and $s = 1/y$. Then as functions on $E_1$, $t$ has a simple zero and $s$ has a zero of order 3 at infinity (Because $x$ is a degree 2 and y is a degree 3 function on $E_1$, [23] example 11.3).

Divide the Weierstrass equation
$$y^2 + a_1 xy + a_3 y = x^3 + a_2 x^2 + a_4 x + a_6$$

through by $y^3$, replacing with $s$ and $t$ and rearranging gives:
$$s = t^3 + -a_1 st + a_2 s^2 - a_3 s^2 + a_4 ts^2 + a_6 s^3.$$

Now substituting the expression for $s$ into the right hand side gives:
$$s = t^3 - a_1(t^3 + -a_1 st + a_2 s^2 - a_3 s^2 + a_4 ts^2 + a_6 s^3)t +$$
$$a_2(t^3 + -a_1 st + a_2 s^2 - a_3 s^2 + a_4 ts^2 + a_6 s^3)t^2 + \cdots.$$

By repeating this substitution until we know the coefficients of $t^3, \cdots, t^9$ gives and expression:

$$\frac{1}{y} = s = t^3(1 - a_1t + (a_1^2 + a_2)t^2 - (a_1^3 + 2a_1a_2 + a_3)t^3 + \cdots)$$

(to see this equality, see [20] section IV.1.) Taking the reciprocal gives an expression for $y$ in terms of $t$:

$$y = t^{-3} + u_1 t^{-2} + u_2 t^{-1} + u_3 + u_4 t + u_5 t^2 + u_6 t^3 + O(t^4)$$

where $O(t^4)$ is a function that has a zero of order 4 at $\infty$ and the coefficients $u_i$ are

$$u_1 = a_1, u_2 = -a_2, u_3 = a_3, u_4 = -(a_1a_3 + a_4),$$
$$u_5 = a_2a_3 + a_1^2 a_3 + a_1 a_4,$$
$$u_6 = -(a_1^2 a_4 + a_1^3 + a_2 a_4 + 2a_1 a_2 a_3 + a_3^2 + a_6).$$

Then the fact that $x = ty$, gives a similar expression for $x$ in terms of $t$:

$$x = t^{-2} + u_1 t^{-1} + u_2 + u_3 t + u_4 t^2 + u_5 t^3 + u_6 t^4 + O(t^5).$$

Then taking $A_1$, $A_2$, $A_3$, $A_4$ and $A_6$ given in the statement of the theorem, and define $F$ to be the function of the Weierstrass equation with these coefficients. Then define $G$ as the function on $E_1$:

$$G = F(\alpha, \beta) = \beta^2 - \alpha^3 + A_1 \alpha\beta - A_2 \alpha^2 + A_3 \beta - A_4 \alpha - A_6.$$

Clearly, $\alpha$ and $\beta$ are rational functions in $x$ and $y$, hence functions on $E_1$. By substituting in the expansions of $x$ and $y$ in $t$ into the formulas for $\alpha$ and $\beta$, an intricate calculation of $F(\alpha, \beta)$ shows that it is $O(t)$. Meaning that the function $G$ vanishes at $\infty$. (This calculation can be done by applying the algorithm for computing the truncated reciprocal in appendix A.2.3.)

The explicit formulas for $\alpha$ and $\beta$ show that they have poles only at points in the set $C$. As $F$ is a polynomial in $\alpha$ and $\beta$, $G$ can only have poles where these functions do. We just saw that $G$ has a zero at $\infty$. It is clear from lemma 3.1.3 that for any $Q \in C$ the rational maps $\alpha$ and $\beta$ are invariant under translation by $Q$, that is $\alpha(P+Q) = \alpha(P)$ and $\beta(P+Q) = \beta(P)$. Thus $G$ is invariant under translation by $Q$ as well. So as $G$ has a zero at $\infty$ then $G$ is zero at every point in $C$. Thus $G$ has no poles, and therefore it must be constant, [23] proposition 11.1. (This is an elementary application of divisors of functions on curves.) It follows that

$$G(P) = F(\alpha(P), \beta(P)) = 0$$

and hence $(\alpha(P), \beta(P))$ satisfies the Weierstrass equation with coefficients $A_i$. Furthermore, $\varphi = (\alpha, \beta)$ preserves the point at infinity, and as mentioned before this implies that $\varphi$ is a group homomorphism. Thus, by definition $\varphi$ is an isogeny.

Now to show that these formulas specify a separable isogeny, we examine the degree of the rational map $\alpha$.

However, to determine this we first define the function $G$ on $E_1$ as follows:

$$G(x, y) = x^3 - y^2 - a_1xy + a_2x^2 - a_3y + a_4x + a_6.$$

From the Weierstrass equation we see that $G$ is 0 for all points on $E_1$. Then by taking the partial derivatives with respect to $x$ and $y$ and evaluating at $Q = (x_Q, y_Q)$ we see:

$$G_x(x_Q, y_Q) = 3x_Q^2 + 2a_2x_Q + a_4 - a_1y_Q = g_Q^x,$$

and

$$G_y(x_Q, y_Q) = -2y_Q - a_1x_Q - a_3 = g_Q^y.$$

Thus if both $g_Q^y$ and $g_Q^x$ are 0, then it is clear that

$$G_x(x_Q, y_Q) = G_y(x_Q, y_Q) = 0$$

and then by example I.1.5 in [20] $Q$ is a singular point, contradicting the fact that $E_1$ is a nonsingular curves. Because $g_Q^y$ is the bivariate non-two torsion polynomial it is 0 if and only if $Q$ is a two torsion point. Therefore, if $Q$ is a two torsion point then $g_Q^x$ cannot be 0. Furthermore, if $Q$ is a two torsion point then $v_Q$ is equal to $g_Q^x$ and hence not 0. Also, if $Q$ is not a two torsion then $u_Q$ is the square of $g_Q^y$ and hence not 0.

For $Q \in S$, we have:

$$\frac{v_Q}{x_P - x_Q} + \frac{u_Q}{(x_P - x_Q)^2} = \frac{v_Q(x_P - x_Q) + u_Q}{(x_P - x_Q)^2}.$$

Let $p_Q(x) = v_Q(x - x_Q) + u_Q$. Then in this notation:

$$\alpha(x) = x + \sum_{Q \in S} \frac{p_Q(x)}{(x - x_Q)^2}.$$

Thus if $Q$ is a two torsion then $u_Q$ is 0 and $v_Q$ is not, hence

$$\frac{p_Q(x)}{(x - x_Q)^2} = \frac{v_Q}{x - x_Q}.$$

Otherwise, if $Q$ is not a two torsion, then $u_Q$ is not 0 and hence

$$\frac{p_Q(x)}{(x - x_Q)^2} = \frac{v_Q(x - x_Q) + u_Q}{(x - x_Q)^2}$$

is in reduced form.

Define $n_Q(x)$ and $d_Q(x)$ to be the numerator and denominator respectively $p_Q(x)/(x - x_Q)^2$ in reduced form. Thus if $Q$ is a two torsion point then $d_Q(x)$ is $x - x_Q$ and otherwise $d_Q(x)$ is $(x - x_Q)^2$. Likewise, if $Q$ is not a two torsion point then $n_Q(x)$ is $v_Q$ and otherwise $n_Q(x)$ is $v_Q(x - x_Q) + u_Q$. Then define the polynomial $\psi$ as follows:

$$\psi(x) = \prod_{Q \in S} d_Q(x).$$

Hence $\psi$ is the denominator of $\alpha$. From the characterization of $d_Q(x)$ it is clear that the highest power of $x - x_Q$ that divides $\psi$ is 1 if $Q$ is a two torsion and 2 otherwise.

Next we define $\psi_Q(x)$ as $\psi(x)/d_Q(x)$, and set

$$r(x) = \sum_{Q \in S} \psi_Q(x) n_Q(x).$$

In the case that $Q$ is a two torsion point then $\deg(\psi_Q)$ is $\deg(\psi) - 1$, and $n_Q$ is constant and hence has degree 0. In the case that $Q$ is not a two torsion point then $\deg(\psi_Q)$ is $\deg(\psi) - 2$, and $n_Q$ is linear and hence has degree 1. Thus in both cases the degree of $\psi_Q(x) n_Q(x)$ is $\deg(\psi) - 1$. So the degree of $r(x)$ is $\deg(\psi) - 1$. Then if we define $p(x)$ as $x\psi(x) + r(x)$ so $\deg(p)$ is $\deg(\psi) + 1$.

Note that the number of points in the set of kernel points $C$ is equal to:

$\#(\text{two torsion points in } S) + 2 \cdot \#(\text{non-two torsion points in } S) + 1.$

This is clear from the way that the set $C$ was partitioned in step 1. Therefore, $\deg(\psi) = \#C - 1$.

Explicitly, the rational map $\alpha$ is of the form:

$$\alpha(x) = x + \frac{r(x)}{\psi(x)} = \frac{p(x)}{\psi(x)}.$$

Then the degree of $p$ is $\#C$, so by definition 2.2.6 the degree of $\varphi$ is $\#C$. Thus by lemma 2.2.14, $\varphi$ is separable.

It is also clear that the leading coefficients of $p$ and $q$ are both 1, so that the coefficient of the pullback of the invariant differential along $\varphi$ must be 1 as well, hence by definition $\varphi$ is normalized.

This concludes the proof that Vélu's formulas define a separable normalized isogeny with specified codomain, and that the isogeny can be computed via the given rational maps $\alpha$ and $\beta$. □

For computational purposes, we would like to know the algebraic complexity of applying Vélu's formulas (for a definition and discussion of algebraic complexity see appendix A.1.) Measuring complexity in this case is somewhat complicated by the fact that while the elliptic curve $E_1$ is defined over $K$, as an algebraic variety it is fully realized over $\overline{K}$ and in many ways only makes sense in this setting. On the one hand, we can look at the algebraic complexity over $\overline{K}$ but this is somewhat unsatisfactory in that our input curve is defined over $K$ and the kernel points are defined over some extension field. Counting $\overline{K}$ operations tells us nothing about the actual time we spend operating over finite precision inputs. What we would really like is to get an idea of the algebraic complexity over $K$. To get the most in-depth look at the algebraic complexity, we need to consider the contribution of both the degree of an isogeny as well as the degree of the extension over which the kernel of the isogeny is defined. (Note that the following theorem uses the *soft-Oh* notation, see definition A.1.5.)

**Theorem 3.5.** *Suppose $\varphi$ is an isogeny of degree $\ell$, where $F/K$ is a minimal degree extension such that the kernel $\varphi$ is contained in $E(F)$. If $d$ is the degree of $F/K$, then running steps 1 through 4 of Vélu's formulas, and also evaluating equations (3.1) and (3.2) on an actual point of $E(K)$ require $O(\ell)$ operations in $\overline{K}$, or $\tilde{O}(\ell M(d))$ operations in $K$.*

*Proof.* The total algebraic complexity can be determined by a step-by-step analysis.

In step 1, partitioning the kernel into lists of two torsion and non-two torsion points requires applying the two division polynomial. For each point in the kernel this requires a constant number of $F$ operations, and checking if the result is zero or not. So for each point in the kernel this is a constant number of algebraic operations, so the total complexity for checking all points in the kernel is $\tilde{O}(\ell)$. Then, sorting the non-two torsion points can be accomplished by ordering the other points by $x$-coordinate, and taking only the even or odd indexed points. This has the well known complexity of $O(\ell log \ell) = \tilde{O}(\ell)$ comparisons (which some authors take to be $F$ operations [5], [8].) Hence this dominates the complexity of this step.

In step 2, for $Q$ in $\ker(\varphi)$ we compute the values $g_Q^x$, $g_Q^y$, $u_Q$ and $v_Q$. If $\ker(\varphi)$ is contained in $E(F)$, the for $Q$ in $\ker(\varphi)$, the associated values $x_Q$ and $y_Q$ are in $F$. Hence, the values $g_Q^x$, $g_Q^y$, $u_Q$ and $v_Q$ are all in $F$ as well, and computing each of them requires a constant number of $F$ operations. The values $v$ and $w$ can be updated at each step, and there are $O(\ell)$ points $Q$ in $S$ so the total complexity is operations in $F$.

In step 3, we compute the Weierstrass model of the domain curve from the values $v$ and $w$ computed in step 2. This only uses a constant number of $F$ operations.

In step 4, we set the rational maps of the coordinates, these are rational maps over $K$ of degree $O(\ell)$.

From this we can see that steps 1 and 2 dominate the algebraic complexity. So that the total complexity for applying Vélu's formulas is $\tilde{O}(\ell)$ operations in $F$. All $F$ operations are $O(M(d))$ operations in $K$, thus we get that computing the total complexity is $\tilde{O}(\ell M(d))$ operations in $K$. Furthermore, once the codomain curve and the rational maps have been computed, these can be stored. Then evaluating the isogeny at a point on the domain curve can then be accomplished in $\tilde{O}(\ell M(d))$ operations in $K$. If we are only counting operations in $\overline{K}$ we can ignore the $M(d)$ factor in the algebraic complexity. □

Ultimately, as in theorem 3.1.5, it is most informative to know exactly what values impact the algebraic complexity. Here we see that the algebraic complexity is primarily dominated by the degree of the extension over which the kernel is defined. However, leaving the complexity analysis in both $\ell$ and $d$ is somewhat unsatisfactory, because it gives the incorrect impression that these two values are independent when they are anything but. Specifically, the extension degree $d$ can be expressed in $\ell$. The following corollary expresses the algebraic complexity of Vélu's formulas uniformly in $\ell$.

**Theorem 3.6.** *For an isogeny $\varphi$ of degree $\ell$ (not divisible by the characteristic of $K$), running Steps 1 through 4 of Vélu's formulas, and also evaluating equations (3.1) and (3.2) on an actual point of $E(K)$ require $\tilde{O}(\ell M(\ell^2))$ operations in $K$.*

*Proof.* It suffices to show that there is an extension $F/K$ such that $\ker(\varphi)$ is contained in $E(F)$ with $[F:K] = O(\ell^2)$.

Because $\ker(\varphi)$ is an order $\ell$ subgroup it is entirely contained in $E[\ell]$, and $E[\ell]$ is isomorphic to $\mathbb{Z}/\ell\mathbb{Z} \times \mathbb{Z}/\ell\mathbb{Z}$ by lemma 2.1.12. So $\ker(\varphi)$ is generated by at most two elements.

If $\ker(\varphi)$ is cyclic let $G = (\alpha, \beta)$ denote a generator. Then $\alpha$ is a root of the square of the $\ell$-torsion polynomial $\psi_\ell^2$ by corollary 2.1.10. By lemma 2.1.7 $\psi_\ell^2$ has degree $\ell^2 - 1$ it follows that $L = K(\alpha)$ is an extension of degree at most $\ell^2 - 1$. Then, $\beta$ is a solution to either a linear or quadratic polynomial over $L$, so that $F = L(\beta)$ is an extension of degree 1 or 2. Thus $G$ in $E(F)$, so the cyclic group generated by $G$, $\ker(\varphi)$, is contained in $E(F)$. Hence $[F:K] = O(\ell^2)$.

Now suppose that $\ker(\varphi)$ is generated by two independent elements $G_1 = (\alpha_1, \beta_1)$ and $G_2 = (\alpha_2, \beta_2)$. (In this case independent means that $\langle G_1 \rangle \cap \langle G_2 \rangle$ is trivial.) If we let $L_1 = K(\alpha_1, \beta_1)$ and $L_2 = K(\alpha_2, \beta_2)$ then $G_1$ and $G_2$ are in $E(L_1)$ and $E(L_2)$ respectively. Thus $E(L_1)$ contains $\langle G_1 \rangle$ and $E(L_2)$ contains $\langle G_2 \rangle$. Now suppose that $\ell_1$ and $\ell_2$ are the orders of $G_1$ and $G_2$ respectively. Then as argued in the cyclic case $[L_1:K] = O(\ell_1^2)$ and $[L_2:K] = O(\ell_2^2)$. Furthermore, $\ker(\varphi)$ is exactly $\langle G_1, G_2 \rangle$, and because $G_1$ and $G_2$ are independent this implies that $\ell = \ell_1 \ell_2$, thus if $F = L_1 L_2$ then $[F:K]$ is $O(\ell_1^2 \ell_2^2) = O(\ell^2)$. Also, both $G_1$ and $G_2$ are in $E(F)$, so $\ker(\varphi)$ is contained in $E(F)$. Hence, in the case that the kernel of $\varphi$ is generated by two independent elements, the points of the kernel are defined over an extension $F/K$ of degree $O(\ell^2)$ just as in the cyclic case. □

Only by considering both theorem 3.1.5 and corollary 3.1.6 does one obtain the most complete view of the complexity of applying Vélu's formulas. One can see by the following example that the minimal degree of the extension can be both 1 and $O(\ell^2)$.

*Example* 3.7. Let $K = \mathbb{F}_7$ and $E_1$ be the elliptic curve defined over $K$ by

$$y^2 = x^3 + x.$$

Likewise, let $E_2$ be the elliptic curve defined over $K$ by

$$y^2 = x^3 - 2.$$

In both cases, the $a$-invariants $a_1$ and $a_3$ are both 0, so the two torsion polynomial $\psi_2(x, y) = 2y$ by definition 2.1.1. Hence the two torsion points are the ones with $y$-coordinate equal to 0.

First we find an isogeny with kernel order $\ell = 2$, such that the kernel is defined over $K$. Thus, the point $P_1 = (0,0)$ on $E_1$ is clearly defined over $K$. So the isogeny $\varphi_1$ with kernel $\langle P_1 \rangle$ has kernel defined over $K$.

Now to observe the opposite extreme, we can show the existence of an isogeny of degree 2 with kernel defined over an extension of degree 3, which is the degree of the (univariate) two torsion polynomial. As the two torsion points on $E_2$ are the points with $y$-coordinate equal to 0, the $x$-coordinate must be a root of the polynomial $x^3 - 2$. Exhaustive search shows that 2 is not a cube modulo 7, so this polynomial is irreducible over $K$. Let $\alpha$ be any cube root of 2 in $\overline{K}$, so $F = K(\alpha)$ is a degree 3 extension, and a minimal degree extension such that the two torsion point $P_2 = (\alpha, 0)$ on $E_2(\overline{K})$ is defined. Thus to define an isogeny $\varphi_2$, of degree $\ell = 2$ with kernel $\langle P_2 \rangle$ it is necessary to work over an extension of degree equal to the degree of the $\ell$-torsion polynomial. As the degree of the $\ell$ is the source of the $O(\ell^2)$ bound in corollary 3.4, this demonstrates an isogeny of degree $\ell$ where kernel is only defined over an extension with degree limited by the degree of the $\ell$-torsion polynomial.

In the worst case scenario, the degree over which the kernel is defined can grow quadratically in $\ell$. However, in practice, the contribution is only dependent on the extension degree over which the kernel is defined, which may be smaller in many other cases.

## 3.2 Kohel's Approach: Computing from the kernel polynomial

In his dissertation, D. Kohel introduced a new approach for determining the domain and rational maps from the kernel of an isogeny [16]. Specifically, as opposed to Vélu's approach of calculating from a list of points in the kernel, Kohel introduced the idea of calculating the isogeny from the kernel polynomial. Specifically, given any finite set $S$ of points on an elliptic curve $E(\overline{K})$, there is a unique monic polynomial of minimal degree, $\psi$, defined over $\overline{K}$ such that $\psi(x) = 0$ if and only if $x$ is the $x$-coordinate of a point in $S$. So, the kernel polynomial of a separable isogeny is the minimal degree polynomial with roots at the $x$-coordinates of the kernel points.

Similar to Vélu's formulas, Kohel's formulas give a straight forward algorithm to calculate the codomain and rational maps of an isogeny. To illustrate this we precisely state the input and output of this algorithm.

**Input:** Given a curve $E_1$ in general Weierstrass form and a kernel polynomial $\psi(x)$ of a separable isogeny. Here, we add the restriction that the kernel associated to $\psi$ is either odd order, or if it is even order that it is contained in or equal to the $E_1[2]$.
**Output:** The general Weierstrass coefficients of a Weierstrass model for the codomain curve $E_2$ of a separable normalized isogeny with kernel polynomial $\psi$. Also, coordinate maps (as rational maps on $E_1$) that evaluate a point $(x, y)$ on $E_1$ to a point on $E_2$.

Given the explicit formulas in this section it is a simple matter to apply them to obtain the algorithm.

As stated we restrict to the cases that the kernel of the isogeny is order two, the whole two torsion, or an odd order. In each case, we assume that the domain of the isogeny is an elliptic curve with Weierstrass model

$$y^2 + a_1xy + a_3y = x^3 + a_2x^2 + a_4x + a_6.$$

Just as in Vélu's formulas, the codomain of the isogeny is given by Weierstrass model:

$$y^2 + a_1xy + a_3y = x^3 + a_2x^2 + (a_4 - 5v)x + (a_6 - (a_1^2 + 4a_2)v - 7w). \quad (17)$$

If the kernel has an odd order $d = 2n + 1$, then the kernel polynomial is given by:

$$\psi(x) = x^n + \sum_{i=0}^{n-1}(-1)^i s_i.$$

If $b_2$, $b_4$, and $b_6$ are the $b$-invariants of the Weierstrass model of the domain, then the codomain of the isogeny is given by equation (3.7) where $v$ and $w$ are given by:

$$v = 6(s_1^2 - 2s_2) + b_2 s_1 + nb_4$$

and

$$w = 10(s_1^3 - 3s_1 s_2 + 3s_3) + 2b_2(s_1^2 - 2s_2) + 3b_4 s_1 + nb_6$$

where $s_2 = 0$ if $n < 2$ and $s_3 = 0$ if $n < 3$.

Then we define the polynomial:

$$\phi(x) = (4x^3 + b_2 x^2 + 2b_4 x + b_6)(\psi'(x)^2 - \psi''(x)\psi(x))$$
$$- (6x^2 + b_2 x + b_4)\psi'(x)\psi(x) + (dx - 2s_1)\psi(x)^2.$$

Then let $\psi_2$ denote the bivariate two torsion polynomial (as in equation (2.2)). If the characteristic of $K$ is not 2, then define:

$$\omega(x,y) = \left(\phi'(x)\psi(x)\frac{\psi_2(x,y)}{2} - \phi(x)\psi'(x)\psi_2(x,y)\right) - \frac{a_1\phi(x) + a_3\psi(x)^2}{2}\psi(x)$$
$$= \left(y + \frac{a_1 x + a_3}{2}\right)(\phi(x)\psi'(x) - \phi'(x)\psi(x)) - \frac{a_1\phi(x) + a_3\psi(x)^2}{2}\psi(x)$$

In the general case, first define:

$$\tilde{\psi} = \sum_{i=0}^{n-2}\binom{i+2}{2}s_{i+2}(-x)^i$$

and

$$\tilde{\tilde{\psi}} = -\sum_{i=0}^{n-3}3\binom{i+3}{3}s_{i+3}(-x)^i.$$

Note, that while similar to derivatives, these polynomials are not exactly the second and third derivative of $\psi$. Then, use these formulas to define:

$$\begin{aligned}\omega(x,y) =& \phi'(x)\psi(x)y - \phi(x)\psi'(x)\psi_2(x,y) + \\ & ((a_1x + a_3)(\psi_2(x,y))^2(\tilde{\psi}(x)\psi'(x) - \tilde{\tilde{\psi}}(x)\psi(x)) + \\ & (a_1\psi_2(x,y)^2 - 3(a_1x + a_3)(6x^2 + b_2x + b_4))\tilde{\psi}(x)\psi(x) + \\ & (a_1x^3 + 3a_3x^2 + (2a_2a_3 - a_1a_4)x + (a_3a_4 - 2a_1a_6))\psi'(x)^2 + \\ & (-(3a_1x^2 + 6a_3x + (-a_1a_4 + 2a_2a_3)) + \\ & (a_1x + a_3)(dx - 2s_1))\psi'(x)\psi(x) + (a_1s_1 + a_3n)\psi(x)^2)\psi(x)\end{aligned}$$

This ends the discussion of the form of the rational maps in the odd degree case.

In the case that the kernel is of order 2, then the kernel polynomial $\psi(x) = x - x_0$ is the single $x$-coordinate of the non-infinite point in the kernel. If $y_0$, the y-coordinate is not known, then it is easily found. If the characteristic of $K$ is 2, then square roots are unique and we can define

$$y_0 = \sqrt{x_0^3 + a_2x_0^2 + a_4x_0 + a_6},$$

otherwise define

$$y_0 = -\frac{a_1x_0 + a_3}{2}.$$

The codomain is given by equation (3.7) where $v$ and $w$ values are given by:

$$v = 3x_0^2 + 2a_1x_0 + a_4 - a_1y_0 \text{ and } w = x_0v.$$

The polynomials $\phi$ and $\omega$ are given by

$$\phi(x) = (x(x - x_0) + v)(x - x_0)$$

and

$$\omega(x,y) = (y(x - x_0)^2 - v(a_1(x - x_0) + (y - y_0)))(x - x_0).$$

In the case that the kernel is the entire two torsion then

$$\psi(x) = x^3 - s_1x^2 + s_2x - s_3$$

and if $b_2$ and $b_4$ are $b$-invariants of the Weierstrass model of the domain, then the codomain is given by the equation (3.7) where the $v$ and $w$ values are given by:

$$v = 3(s_1^2 - 2s_2) + \frac{b_2s_1 + 3b_4}{2} \text{ and } w = 3(s_1^3 - 3s_1s_2 + 3s_3) + \frac{b_2(s_1^2 - 2s_2) + b_4s_1}{2}.$$

(Note that we are assuming the characteristic of $K$ is not 2 here, this is fine because as we saw in section 2.1 multiplication by 2 is not separable in such fields.) Then define the polynomials:

$$\phi_1(x) = \psi'(x)^2 + (-2\psi''(x) + (4x - s_1))\psi(x),$$

and
$$\omega_1(x,y) = \frac{\psi_2(x,y)(\phi_1'(x)\psi(x)\phi_1(x)\psi'(x)) - (a_1\phi_1(x) + a_3\psi(x))\psi(x)}{2}.$$

Then we set
$$\phi(x) = \phi_1(x)\psi(x) \text{ and } \omega(x,y) = \omega_1(x,y)\psi(x).$$

This concludes the characterization of the rational maps in the case that the kernel is contained in $E_1[2]$.

We summarize the results of this section as:

**Theorem 3.8.** *If $\psi$ is the kernel polynomial of an isogeny of odd degree, degree 2, or degree 4 if the kernel is also the entire two torsion. The codomain of the isogeny is given by equation (3.7). The rational maps for the isogeny are given by:*
$$\left(\frac{\phi(x)}{\psi(x)^2}, \frac{\omega(x,y)}{\psi(x)^3}\right),$$
*where $\psi$, $\phi$, $\omega$, $v$ and $w$ are given for the individual cases above.*

We state this without proof, because it will follow from the results of the next section.

Certainly Kohel's formulas simplify the task of computing an isogeny. Indeed these formulas reduce the problem of computing the isogeny to polynomial arithmetic and evaluation. However, it is not immediately clear why this is at all useful, other than conceptual clarity. It does, in fact, work out that Kohel's formulas do provide a performance improvement, in some cases.

Specifically, the kernel of an isogeny is defined over some algebraic extension $L/K$, it may occur that the kernel polynomial of an isogeny $\psi(x)$ is defined over an intermediate extension $F$ properly contained in $L$. When the field of definition of $\psi$ is of lower degree than the field of definition of the actual points of the kernel, this can lead to a speed up by simplifying the necessary extension field arithmetic.

Immediately the question comes to mind: Does it ever occur that using Kohel's formulas provides a speed up? Alternately, does it ever occur that using Kohel's formulas does not provide a speed up over Vélu's formulas? The answer to both questions is yes. This can be seen by examining two extreme cases: When $\psi$ is defined over $K$ or when the kernel polynomial is only defined over $L$, the extension of definition of the kernel of the isogeny. In the first case, it is sufficient to only perform polynomial arithmetic over $K$. However, in the second case Kohel's algorithms will not provide any speed up over Vélu's formulas. The following example illustrates how each of these cases may be realized.

*Example* 3.9. Consider the elliptic curve $E$ defined over $K = \mathbb{F}_7$ by the Weierstrass equation $y^2 = x^3 + x + 1$. In this case the $a$-invariants $a_1$ and $a_3$ are both zero, so that the two torsion points on $E(\overline{K})$ are the points with $y$-coordinate 0.

Thus, we see that the minimal polynomial over $K$ of the $x$-coordinates of the two torsion points is $\psi_2(x) = x^3 + x + 1$. We can easily check that this polynomial is irreducible, by checking that it has no roots in $K$ (as it is degree 3, if it were reducible, it must have a linear factor.) We can further note that this polynomial is separable (i.e. has no repeated roots) because $\psi'(x) = 3x^2 + 1 = 3(x^2 + 5)$ and we can tell that if it did $\psi(x)$ would have a nontrivial factor, contradicting the fact that it is irreducible.

Now consider a root $\alpha$ of $\psi_2$ and let $F = K(\alpha)$. It turns out that $\psi_2$ splits completely over $F$, with roots $\alpha$, $2\alpha^2 + 6$ and $-2\alpha^2 + 4\alpha + 2$. So that the nontrivial two torsion points are:

$$(\alpha, 0), (2\alpha^2 + 6, 0) \text{ and } (-2\alpha^2 + 4\alpha + 2, 0)$$

and all are contained in $E(F)$.

Suppose that $\varphi_1$ is the isogeny with the full two-torsion set as kernel, then applying Vélu's formulas requires working over $F$. However, as the kernel polynomial is defined over $K$, applying Kohel's formulas requires that we work over $K$. This realizes the case when the kernel polynomial is defined over $K$ but the kernel is defined over an extension.

However, we can also see the other extreme case, when the kernel polynomial is only defined over the same extension field as the points of the kernel. If $\varphi_2$ is the isogeny with order 2 kernel $\langle(\alpha, 0)\rangle$, applying Vélu's formulas requires that we work over $F$. However, the kernel polynomial $\psi$ of $\varphi$ is the linear polynomial $x - \alpha$. This polynomial is defined over $F$, so applying Kohel's formulas requires working over $F$ as well.

More formally, let $\psi(x)$ in $\overline{K}[x]$ be the kernel polynomial of the degree $\ell$ separable isogeny $\varphi$. Let $F/K$ be an algebraic extension of minimal degree $d$ such that the coefficients of $\psi(x)$ are all contained in $F$. Then because extension field arithmetic is implemented via polynomial arithmetic, one $F$ multiply takes $M(d)$ operations in $K$, so that we can take all $F$ operations to be $O(M(d))$ operations in $K$. As multiplication is the limiting factor in extension field arithmetic we will take all $F$ operations to be $O(M(d))$ operations in $K$. The polynomial $\psi$ has degree $\ell$, so that the degree of all polynomials involved is $O(\ell)$. Thus, the polynomial arithmetic in $F[x]$ requires $O(M(\ell))$ operations in $F$, or $\tilde{O}(M(d\ell))$ operations in $K$.

This result is summarized in the following theorem:

**Theorem 3.10.** *If $\varphi : E_1 \to E_2$ is an isogeny with kernel polynomial $\psi$ in $F[x]$, where $F/K$ is an algebraic extension of degree $d$, then Kohel's formulas can be computed in $\tilde{O}(M(d\ell))$ operations in $K$. The formulas can be precomputed and then evaluated in $O(\ell M(d))$ operations in $K$.*

So if we restrict to the case of an isogeny defined over $K$, then the kernel polynomial must be defined over $K$ as well. In this case $F = K$ and hence $d = 1$ so that we get the following corollary.

**Corollary 3.11.** *If $\varphi : E_1 \to E_2$ is an isogeny defined over $K$ then Kohel's formulas can be computed in $O(M(\ell))$ operations in $K$. The formulas can be precomputed and then evaluated in $O(\ell)$ operations in $K$.*

It is common to consider only isogenies defined over $K$, such as in the case of the SEA point counting algorithm ([2], [1] chapter VII.) So the assumptions of corollary 3.1.11 are not unreasonable in practice.

# 4 Computing from the kernel polynomial: General degree isogenies

Kohel's idea is a very useful observation, and in many cases leads to improved performance due to performing the computations over a lower degree extension field. However, in some cases the restriction of working over an odd degree field may be overly restrictive. This idea can be generalized to work over arbitrary degree isogenies [2].

In order to compute a general degree isogeny from a kernel polynomial, we stipulate that the domain curve must be in short Weierstrass form. This greatly simplifies the algebra.

**Theorem 4.1.** *Suppose $E_1$ is an elliptic curve in short Weierstrass form*

$$y^2 = x^3 + Ax + B.$$

*Let $\psi$ be the kernel polynomial of a separable normalised isogeny $\varphi$ with domain $E_1$ and degree $\ell$. Let $\psi_2 = \gcd(x^3 + Ax + B, \psi)$. Then define*

$$D(x) = \psi^2/\psi_2 = x^{\ell-1} - \sigma_1 x^{\ell-2} + \sigma_2 x^{\ell-3} - \sigma_3 x^{\ell-3} + \cdots.$$

*Then the coordinate maps of $\varphi$ are given by*

$$\alpha(x) = \ell x - \sigma_1 - (3x^2 + A)I(x) - 2(x^3 + Ax + B)I'(x),$$

*where $I(x) = \frac{D'(x)}{D(x)}$, and*

$$\beta(x, y) = y\alpha'(x).$$

*And the codomain curve $E_2$ is given by*

$$y^2 = x^3 + (A - 5v)x + (B - 7w),$$

*where $v = A(\ell-1)+3(\sigma_1^2 - 2\sigma_2)$ and $w = 3A\sigma_1 + 2B(\ell-1) + 5(\sigma_1^3 - 3\sigma_1\sigma_2 + 3\sigma_3)$.*

*Proof.* The proof of this fact follows from the proof of Vélu's formulas. As in Vélu's formulas $C$ is the set of points of the kernel of a separable normalized isogeny $\varphi$.

First we determine the map $\alpha$. Then by lemma 3.1.3 $\alpha(P)$ is given by

$$x_P + \sum_{Q \in C} x_{P+Q} - x_Q.$$

As in Vélu's formulas (section 3.1.1) $C$ is partitioned into the disjoint sets $\{\infty\}$, $C_2$, $R$ and $-R$ where $C_2$ are the two torsion points and $R$ and $-R$ are the rest of the points of $C$ sorted from their inverses. Thus we can write $D(x)$ as:

$$\prod_{Q \in C_2} (x - x_Q) \prod_{Q \in R} (x - x_Q)^2.$$

Thus if we let $C^+$ be $C_2 \cup R$. Then by equations (3.3) and (3.5) in lemma 3.1.3, if $P$ is $(x, y)$ then

$$\alpha(P) = x + \sum_{Q \in C^+} \left( \frac{v_Q}{x - x_Q} + 4 \frac{x_Q^3 + Ax_Q + B}{(x - x_Q)^2} \right)$$

where $v_Q$ is $3x_Q^2 + A$ if $Q$ is a two torsion and $v_Q$ is $2(3x_Q^2 + A)$. As $E_1$ is in short Weierstrass form, when $Q$ is a two torsion $x_Q^3 + Ax_Q + B$ is 0. Hence

$$\sum_{Q \in C_2} x_{P+Q} - x_Q = \sum_{Q \in C_2} \left( \frac{v_Q}{x - x_Q} + 2 \frac{x_Q^3 + Ax_Q + B}{(x - x_Q)^2} \right)$$

Now looking at a similar sum for the non-two torsion points gives that the sum over $Q \in R \cup -R$ is

$$\frac{1}{2} \sum_{Q \in C - C_2} \left( \frac{v_Q}{x - x_Q} + 4 \frac{x_Q^3 + Ax_Q + B}{(x - x_Q)^2} \right).$$

Combining these two equations gives:

$$\alpha(P) = \sum_{Q \in C - \{\infty\}} \left( \frac{3x_Q^2 + A}{x - x_Q} + 2 \frac{x_Q^3 + Ax_Q + B}{(x - x_Q)^2} \right).$$

It is straight forward algebraic manipulation to show

$$x - x_Q - \frac{3x^2 + A}{x - x_Q} + 2 \frac{x^3 + Ax + B}{(x - x_Q)^2} = \frac{3x_Q^2 + A}{x - x_Q} + 2 \frac{x_Q^3 + Ax_Q + B}{(x - x_Q)^2}.$$

substituting this into the expression for $\alpha$ gives

$$\alpha(P) = \ell x - \sigma_1 - (3x^2 + A) \sum_{Q \in C - \{\infty\}} (x - x_Q)^{-1} + 2(x^3 + Ax + B) \sum_{Q \in C - \{\infty\}} (x - x_Q)^{-2}.$$

The sum of $(x - x_Q)^{-1}$ is $I(x) = D'(x)/D(x)$. The derivative of $(x - x_Q)^{-1}$ is $-(x - x_Q)^{-2}$ so the second sum is equal to $-I'(x)$. This gives the statement of the Theorem.

Given that the $x$-coordinate map is $\alpha$ and $E_1$ is in short Weierstrass form, then the expression for $\beta$ comes from lemma 2.2.21.

The expression for the coefficients of the codomain curves are given by expanding the values from Vélu's formulas $A - 5v$ and $B - 7w$ in the symmetric functions (coefficients) of $D(x)$. Hence the expressions in $\sigma_1$, $\sigma_2$ and $\sigma_3$. □

From this theorem, it seems that to obtain a formula that works for general degree isogenies, we have sacrificed generality of the curves that we consider. Indeed we restrict ourselves to the case of curves in short Weierstrass form. On the other hand, if the characteristic of $K$ is not 2 or 3, then all curves are isomorphic to a curve in short Weierstrass form. This is certainly encouraging, and turns out to be quite useful. However, Vélu's and Kohel's formulas for computing isogenies are quite dependent on the underlying Weierstrass model. We would like to precisely describe how to use theorem 3.2.1 to actually calculate the rational maps and codomain curve given a kernel polynomial of an isogeny with domain in general Weierstrass form. The only restriction we make is that $K$ must be of characteristic not 2 or 3.

First we note that in both Vélu's and Kohel's formulas, the codomain of the computed isogeny has the same coefficients on the $y$, $xy$, and $x^2$ coefficients as the domain curve. In the algorithm that we give, we maintain the same convention. However, the argument that we present will show that one can easily post compose with any Weierstrass isomorphism to obtain a separable isogeny. The approach is straight forward, but requires that we be careful when we pre-compose the isogeny with a Weierstrass isomorphism.

**Input:** A curve $E_1$ in general Weierstrass form, defined over a curve of characteristic not 2 or 3. A kernel polynomial $\psi$ of a separable normalized isogeny $\varphi$ with domain $E_1$ and degree $\ell$.

**Output:** The general Weierstrass coefficients of a Weierstrass model for the codomain curve $E_2$ (with the same coefficients on $y$, $xy$ and $x^2$ as $E_1$) of a separable normalized isogeny with kernel polynomial $\psi$. Also, coordinate maps (as rational maps on $E_1$) that evaluate a point $(x, y)$ on $E_1$ to a point on $E_2$.

1. Calculate $s = -a_1/2$, $r = -(a_2 - sa_1 - s^2)/3$ and $t = -(a_3 + ra_1)/2$. Then define the Weierstrass isomorphism $\rho : E_1 \to \tilde{E}_1$ by

$$\tilde{x} = x - r, \quad \tilde{y} = y - sx + rs - t$$

and its inverse $\rho^{-1} : \tilde{E}_1 \to E_1$ is given by

$$x = \tilde{x} + r, \quad y = \tilde{y} + s\tilde{x} + t.$$

Then $\tilde{E}_1$ is in short Weierstrass form with coefficients

$$A = a_4 - sa_3 + 2ra_2 - (t + rs)a_1 + 3r^2 - 2st$$

and

$$B = a_6 + ra_4 + r^2 a_2 + r^3 - ta_3 - t^2 - rta_1.$$

2. Define $\tilde{\psi} = \psi \circ \rho^{-1}$. Use theorem 3.2.1 with domain curve $\tilde{E}_1$ and kernel polynomial $\tilde{\psi}$ to calculate an isogeny $\tilde{\varphi} : \tilde{E}_1 \to \tilde{E}_2$.

3. Let $A_2$ and $B_2$ be the coefficients of $\tilde{E}_2$. Let $r$, $s$, and $t$ be as in the first step. Define the Weierstrass isomorphism $\tau : \tilde{E}_2 \to E_2$ by

$$x' = \tilde{x} + r, \quad y' = \tilde{y} + s\tilde{x} + t.$$

Then $E_2$ has coefficients $a_1' = a_1$, $a_2' = a_2$, $a_3' = a_3$,

$$a_4' = A_2 + sa_3 - 2ra_2 + (t+rs)a_1 - 3r^2 + 2st$$

and

$$a_6' = B_2 - ra_4 - r^2 a_2 - r^3 + ta_3 + t^2 + rta_1.$$

4. Calculate $\varphi = \tau \circ \tilde{\varphi} \circ \rho$.

**Corollary 4.2.** *This algorithm correctly returns a separable normalized isogeny of degree $\ell$ with codomain $E_1$ and kernel polynomial $\psi$.*

*Proof.* This is succinctly summarized in the following commutative diagram:

$$\begin{array}{ccc} E_1 & \xrightarrow{\varphi} & E_2 \\ \rho \downarrow & & \uparrow \tau \\ \tilde{E}_1 & \xrightarrow{\tilde{\varphi}} & \tilde{E}_2 \end{array}$$

The equations for $\rho$, $\rho^{-1}$ and the curve $\tilde{E}_1$ are correct based on the properties of Weierstrass isomorphisms ([20] III.1.2.) The isogeny $\tilde{\varphi}$ is separable and has codomain $\tilde{E}_2$ by theorem 3.2.1. Once again, the Weierstrass isogeny $\tau$ and codomain $E_2$ are also correct based on the properties of Weierstrass isomorphisms. The only tricky thing is to note is that $\tilde{\psi} = \psi \circ \rho^{-1}$ is the kernel polynomial of $\tau \circ \tilde{\varphi}$. Then, because $\rho^{-1} \circ \rho$ is the identity, it follows that $\psi$ is the kernel polynomial of $\varphi$.

The composite map $\varphi$ is a separable isogeny, as $\rho$, $\tilde{\varphi}$ and $\tau$ are separable isogenies (Weierstrass isomorphisms are degree 1 isogenies.)

It also follows that $\varphi$ is normalized, as $\tilde{\varphi}$ is normalized and because $\rho$ and $\tau$ have no scaling factors. Thus the pullback of the invariant differential of $E_1$ along the composite map has no scaling factors introduced. $\square$

We conclude this section with a brief discussion of the algebraic complexity of applying the algorithms of this section. Unsurprisingly the complexity of this algorithm is not terribly different than applying Kohel's formulas. Unless one calculates out the composite rational map as a quotient of polynomials written out as the canonical sum of multiples of powers of $x$, In which case, the complexity can gain a factor of $\ell$.

**Theorem 4.3.** *Let $\varphi : E_1 \to E_2$ be an isogeny of degree $\ell$ with kernel polynomial $\psi \in F[x]$, where $F$ is some degree $d$ algebraic extension of $K$. Then computing $\varphi$ by the algorithm of this section takes $O(M(\ell d))$ operations in $K$.*

*If one leaves $\varphi$ as a sequence of maps (instead of computing the explicit composite) the complexity of applying these formulas is $O(\ell M(d))$ If one computes out the rational maps for $\varphi$ as a quotient of polynomials, written as a sum of multiples of powers of x then this step takes $O(M(d\ell)\sqrt{\ell})$ operations in K and will dominate the complexity of precomputing $\varphi$.*

*Proof.* The dominant factor in applying the formulas of theorem 3.2.1 is polynomial multiplication. These polynomials are of degree $O(\ell)$ over $F$ hence the dominant algebraic complexity is $O(M(d\ell))$. Computing the Weierstrass isomorphisms takes a constant number of $F$ operations, and hence does not contribute to the algebraic complexity.

Likewise evaluating the Weierstrass isomorphisms takes a constant number of $F$ operations. However, as in the case of Kohel's formulas, this takes $O(\ell M(d))$ operations in $K$.

There are multiple algorithms for evaluating the composite $\tilde{\varphi} \circ \rho$, the best asymptotic complexity that does not contain a dependence on the underlying field is
$$O(M(\ell)\sqrt{\ell \log \ell}) = O(M(\ell)\sqrt{\ell})$$
operations in $F$ ([2] 2.5, [3].) Hence this is $O(M(\ell d)\sqrt{\ell})$ $K$ operations. □

## 5 Computing from Domain and Codomain

The algorithms from the previous section show how to determine the coordinate maps and codomain of an isogeny given a domain and kernel. However, there is a sort of inverse question to this. Suppose we have the domain and codomain of a degree $\ell$ separable isogeny. Can we recover the kernel of this isogeny? Fortunately, the answer is yes.

In this section we prove this by displaying a naive algorithm to recover the kernel, given a domain and codomain. However, this naive algorithm has abysmal performance, so we also present Stark's algorithm which can achieve a much better complexity with a few assumptions about the input.

### 5.1 A Naive Approach

Here we briefly sketch a brute force approach for recovering the kernel of an isogeny from the domain and codomain. So given a domain $E_1$, a codomain $E_2$ and a degree $\ell$. We will only suppose that $p = \text{char}(K)$ does not divide $\ell$. Then we search for the kernel of $\varphi : E_1 \to E_2$ as follows. As $\ker(\varphi)$ is of order $\ell$, it is contained in $E_1[\ell]$, the $\ell$ torsion of $E_1$. By lemma 2.1.12 $E_1[\ell]$ is isomorphic to $\mathbb{Z}/\ell\mathbb{Z} \times \mathbb{Z}/\ell\mathbb{Z}$. So we enumerate all $\ell$ order subgroups $S$ of $E_1[\ell]$ and run Vélu's formulas on each one, checking if the calculated codomain is isomorphic to $E_2$. If we find one, then there is a separable isogeny $\varphi : E_1 \to E_2$ with kernel $S$. If we do not find any such kernel, then there is not a degree $\ell$ isomorphism from $E_1$ to $E_2$.

Next, we briefly analyze the algebraic complexity of this approach when $\ell$ is prime. For $\ell$ prime the group $\mathbb{Z}/\ell\mathbb{Z} \times \mathbb{Z}/\ell\mathbb{Z}$ has $\ell+1$ subgroups of order $\ell$ and in this case Vélu's formulas has algebraic complexity $O(\ell M(\ell^2))$. So this algorithm has algebraic complexity $O(\ell^2 M(\ell^2))$. In general, if $\ell$ is composite then $E[\ell]$ has more than $\ell+1$ order $\ell$ subgroups, so this complexity can not be better than $O(\ell^2 M(\ell^2))$. This complexity is worse than $O(\ell^4)$ and hence this is not a particularly practical algorithm.

## 5.2 Stark's Algorithm

In contrast to the Naive approach in the previous section, Stark's algorithm is a subcubic algorithm for computing the kernel polynomial of an isogeny given the degree, domain and codomain. The main idea underlying this algorithm is that if there exists an isogeny $\varphi : E_1 \to E_2$ with $x$-coordinate map $N(x)/D(x)$ and $\wp_1$ and $\wp_2$ are the respective Weierstrass functions of $E_1$ and $E_2$ then $\wp_1$ and $\wp_2$ are related by

$$\wp_2(z) = \wp_1\left(\frac{N(z)}{D(z)}\right)$$

([2] 6.1.) In [21] Stark proposed a continued fraction approach to recover the rational function $N(z)/D(z)$. Specifically by expanding $\wp_2$ as a continued fraction in $\wp_1$, hence approximating $N(z)/D(z)$. This algorithm has been written up in [2], a more clearly written version of the algorithm occurs in Moody's dissertation ([18] algorithm 3.) The algorithm operates as follows:

**Input:** Given a domain $E_1$ and codomain $E_2$ both in short Weierstrass form of a degree $\ell$ isogeny $\varphi$, where $4\ell < p$, in the case of positive characteristic $p$.
**Output:** The denominator $D(x)$ of the $x$-coordinate map of $\varphi$.

1. Let $S = \wp_1 \mod z^{4\ell}$
2. Let $T = \wp_2 \mod z^{4\ell}$
3. Set $n = -1$, $q_{-2} = 1$, and $q_{-1} = 0$.
4. While $\deg(q_n) < \ell - 1$ do:
   (a) Find $r$ and $t_{-2r}$ such that
   $$T(z) = \frac{t_{-2r}}{z^{2r}} + \cdots + t_0 + t_2 z^2 + \cdots.$$
   (b) Set $n = n+1$ and $a_n = 0$.
   (c) While $0 \leq r$ do:
      i. Set $a_n = a_n + t_{2r} z^r$
      ii. Set $T = T - t_{-2r} S^r \mod z^{4\ell}$.

iii. Find $r$ and $t_{-2r}$ such that

$$T(z) = \frac{t_{-2r}}{z^{2r}} + \cdots + t_0 + t_2 z^2 + \cdots.$$

(d) Set $q_n = a_n q_{n-1} + q_{n-2}$.

(e) If $n = \ell - 1$ go to step 5.

(f) Set $T(z) = 1/T(z) \mod z^{4\ell}$.

5. Return $D(x) = q_n(x)$.

This algorithm is straight forward, except for a few steps. In step 4f computing the truncated reciprocal can be done in time $O(M(\ell))$ by the algorithm stated in section A.2.3. So that the complexity of the main loop is $O(\ell M(\ell))$ Also, we have not shown how to compute $\wp_1$ and $\wp_2$ as in steps 1 and 2, which we will proceed to show.

Before moving on to discussing how to compute the Weierstrass functions on $E_1$ and $E_2$, we make some general remarks about this algorithm.

*Remark* 5.1. This algorithm assumes that the input is in short Weierstrass form. As we will show, this is a requirement of the algorithm for computing $\wp_1$ and $\wp_2$. However, using the methods described in section 3.2 for characteristic not 2 or 3 we can calculate $\tilde{E}_1$ and $\tilde{E}_2$ in short Weierstrass form, and isomorphic to to $E_1$ and $E_2$ respectively. Then by appropriately pre and post composing with these isomorphisms we can determine the isogeny $\varphi : E_1 \to E_2$. Also, note that restricting to characteristic 2 and 3 is implied by the fact that $4\ell < p$ in the case of positive characteristic $p$.

*Remark* 5.2. Notice that this algorithm outputs the denominator $D(x)$ of $\varphi_x$. However the algorithms in sections 3.1.2 and 3.2 take input as a kernel polynomial $\psi$. However, $\psi$ and $D$ are simply related as $\psi = \psi_2 \psi_{>2}$, where $\psi_2$ is the greatest common divisor of $\psi$ and the univariate two torsion polynomial of $E_1$. Then $D = \psi_2(\psi_{>2})^2$. Thus we can easily compute $\psi$ from $D$.

*Remark* 5.3. Note that the value $T$ in this algorithm is always a Laurent series in $z^2$. From an implementation point of view, it is straight forward to store this as a Laurent series in $z$. However this requires storing twice as many coefficients, half of which will always be 0, and will waste operations while performing operations on $T$. By careful implementation, one can succinctly store this Laurent series and perform operations on it that do not waste cycles performing multiplications by 0.

The remainder of this section is on how to compute the Weierstrass $\wp$ function of a curve in short Weierstrass form. We do not go into any background details of this function and point the interested reader to [20] VI.3.3. For our purposes, the $\wp$ function is a Laurent series over $K$ of the form

$$\wp(z) = \frac{1}{z^2} + \sum_{i=0}^{\infty} c_i z^{-2i}. \tag{18}$$

Furthermore, $\wp$ satisfies the differential equation
$$(\wp'(z))^2 = 4\left(\wp(z)^3 + A\wp(z) + B\right) \tag{19}$$

We now give two approaches for solving for $\wp \mod z^n$. First we give a straight forward algorithm with complexity $O(n^2)$ and then give an algorithm with complexity $O(M(n))$.

The first straight forward approach ([2] 3.2) is to combine equations (3.8) and (3.9) and use the fact that
$$c_1 = -\frac{A}{5}, \text{ and } c_2 = -\frac{B}{7}.$$

Differentiating equation (3.9) gives
$$\wp''(z) = 6\wp(z)^2 + 2A.$$

Then solving for $c_j$ gives
$$c_j = \frac{3}{(j-2)(2j+3)} \sum_{i=1}^{j-2} c_i c_{j-1-i}.$$

Directly computing these coefficients then takes $O(n^2)$ operations in $K$. Note that this implies that if we are working in positive characteristic $p$, then we must have $2n + 3 < p$, otherwise the formula will have a division by 0.

The second approach is a more complicated algorithm, but it can solve for $\wp$ in time $O(M(n))$. This approach was introduced in [2] 3.3, and proceeds as follows. Let
$$Q(z) = \frac{1}{\wp(z)} \text{ and } R(z) = \sqrt{Q(z)}$$
where either choice of square root will do here. Then
$$R'(z)^2 = BR(z)^6 + AR(z)^4 + 1.$$

Thus calculating out the first 3 terms of $R$ gives
$$R(z) = z + \frac{A}{10}z^5 + \frac{B}{14}z^7 + \cdots$$

squaring implies that
$$Q(z) = z^2 + \frac{A}{5}z^6 + \frac{B}{7}z^8 + \cdots .$$

This in turn implies that
$$\wp(z) = \frac{1}{z^2} - \frac{A}{5}z^2 + \frac{B}{7}z^4 + \cdots .$$

So this yields the algorithm to compute $\wp(z)$ as follows:
**Input:** $A$ and $B$ coefficients of an elliptic curve $E$ in short Weierstrass form, and degree $n$.
**Output:** The truncated Weierstrass function $\wp \mod z^n$ associated to $E$.

1. Compute $R \mod z^{2n+6}$ by the algorithm for first order nonlinear differential equations in section A.2.2 with $G(t) = Bt^6 + At^4 + 1$.

2. Compute $Q = R \mod z^{2n+5}$.

3. Compute $\wp = 1/Q \mod z^{2n+1}$.

The algorithm for solving first order nonlinear differential equations in section A.2.2 requires $O(M(n))$, as does squaring and reciprocal (see appendix A.2.3) so that the total complexity for computing $\wp$ is $O(M(n))$.

Because computing the functions $\wp_1$ and $\wp_2$ have complexity $O(M(n))$ these steps do not impact the algebraic complexity of Stark's Algorithm. Thus we can take the algebraic complexity of Stark's algorithm as $O(nM(n))$.

# Part IV
# Algebraic Complexity Theory and Algorithms

This appendix provides some background on algebraic complexity theory and some efficient polynomial arithmetic algorithms. These two subjects are very broad in their own right and the purpose of this appendix is not to provide any sort of deep or complete introduction. Both subjects also provide powerful tools for analyzing and understanding number theoretic algorithms. The computational aspects of elliptic curves and isogenies are no exception, so this material is useful for deeper understanding of the results in the body of this document. The results collected here are provided as either a refresher or brief introduction to the few results that are used through out the rest of this document. There are several very good introductions to this deep subject such as [5] and [8].

# A  Algebraic Complexity

Simply put, the algebraic complexity of a number theoretic algorithm measures the number of mathematical operations that the algorithm takes, and how this value scales with the input size. In the case of combinatorial algorithms, the analysis is based on how many operations they use based on the number of bits of input ([11] I.1.) While in many cases this closely corresponds to the algebraic complexity, measuring the number of operations of a mathematical algorithm based on the number of bits in the input may not be particularly informative. For instance, using the number of bits in the input presupposes a fixed representation, and constantly considering this may be cumbersome. To be more precise, algebraic complexity is measured as follows, for a given algorithm, the size of the input is measured in one or more variables that captures the size of the input. Then for a given input, the underlying ring or field of the input is identified. The algebraic complexity is measured as how many operations in the underlying field or ring are used based on the size of the input. Ring operations are considered as addition, subtraction, and multiplication. Field operations are the ring operations as well as inversion. Sometimes, (depending on the context) comparison operations such as equals, greater than or less than are considered ring/field operations as well.

This definition may seem somewhat uninformative, so it is useful to look at a few examples.

*Example* A.1. Consider polynomial arithmetic: given two polynomials $f$ and $g$ in $K[x]$ for some field $K$ with $n = \min\{\deg(f), deg(g)\}$. The complexity of the polynomial arithmetic is measured as the number of $K$ operations it takes to compute $h = f + g$, $h = f - g$ or $h = f \cdot g$ respectively. Specifically, addition and subtraction both take $n$ $K$ operations, measuring multiplication is more complicated and we denote the number of operations as $M(n)$.

*Example* A.2. Consider matrix addition and multiplication of two $m \times m$ matrices over a ring $R$, we consider the size of the input to be the dimension $m$. In this case the naive algorithm for addition requires exactly $m^2$ additions. The naive algorithm for multiplications requires computing $m^2$ inner products, each of which takes $m$ multiplications in $R$ and $m-1$ additions in $R$. Thus the naive algorithm for multiplication takes $2m^3 - m^2$ operations in $R$.

Ultimately, it would be nice to not worry about the exact number of algebraic operations an algorithm uses, but rather to just get an idea of the way that this number scales as the input size changes. The solution is to measure the *asymptotic* complexity of the algorithm, that is if the input grows arbitrarily large, we want a function that dominates the number of operations the algorithm takes. This can be rigorously defined ([8] 25.7), but we first must introduce the following definition:

**Definition A.3.** A function $f : \mathbb{N} \to \mathbb{R}$ is *eventually positive* if there exists some $N$ such that for all $n \geq N$, $f(n)$ is positive.

The asymptotic complexity of an algorithm is measured in *big-oh* notation. The precise definition and notation of this is:

**Definition A.4.** For an input of size $n$ the number of underlying field (ring) operations used by an algorithm is $O(f(n))$, if $f$ is an eventually positive function, and there exists a positive integer $N$ such that for all $n \geq N$, there exists a positive constant $C$ such that for input size of $n$, the algorithm uses no more than $Cf(n)$ operations.

From this it is clear that the naive algorithm for matrix multiplication in example A.1.2 is $O(m^3)$. Which would have been a much easier analysis than actually counting each operation.

The notion of big-oh notation can be relaxed to ignore logarithmic factors, and his is called *soft-oh* notation. The precise definition of this is as follows:

**Definition A.5.** For an input of size $n$ the number of underlying field (ring) operations used by an algorithm is $\tilde{O}(f(n))$, if $f$ is an eventually positive function, and there exists a positive integer $N$ such that for all $n \geq N$, there exists positive constants $b$ and $C$ such that for input size $n$, the algorithm uses no more than $C(\log(3 + f(n)))^b f(n)$ operations.

To illustrate the differences between these two asymptotic measurements, consider the algebraic complexity of matrix multiplication in example A.1.2. Because any algorithm must keep track of the indices of the elements of the matrices that are being multiplied, this requires keeping around counters that can hold values up to $m$ so the length of these variables (in bits) and complexity of arithmetic is $O(\log m)$. If we take these operations into account the algorithm has complexity $O(m^3 \log m)$. But this is cumbersome, and uninformative because as $m$ grows the $m^3$ term will dominate the $\log m$ factor, so it is convenient to consider the soft-Oh asymptotic complexity $\tilde{O}(m^3)$.

*Remark* A.6. There are algorithms that are, in practice, better than the naive algorithm for matrix multiplication. In this case, computing the product of two $m \times m$ matrices is $\tilde{O}(m^\omega)$ where $\omega$ can be taken to be at most $\log_2 7 = 2.807...$ ([8] 12.1.)

# B    Efficient Polynomial Arithmetic

As mentioned in the previous section the algebraic complexity of polynomial multiplication is more complicated than just the naive algorithm that uses $O(n^2)$ operations in the underlying field $K$ (where $n$ is the degree of the polynomials.) For example, the algorithm for fast Fourier multiplication has algebraic complexity $O(n \log n \log \log n) = \tilde{O}(n)$ ([8] 8.2.) Although, the values for $n$ where the exact run times are better than other methods may be quite high. This is just an indicator of how complicated analyzing the costs of polynomial multiplication can be. Herein, we are not interested in the exact complexity of polynomial multiplication. As such, we treat polynomial multiplication as a subroutine and denote the cost as $M(n)$. However, we do make the assumption that the complexity of polynomial multiplication is "superlinear," and by this mean that

$$\frac{M(m)}{m} \leq \frac{M(n)}{n}$$

when $m \leq n$. This implies that

$$\sum_{i=1}^{i} M(2^i) \leq 2M(2^i), \tag{20}$$

a fact that will come in handy when analyzing algorithms for computing truncated power series of polynomials [2].

Recall that Starks algorithm (section 3.3.2) requires that we compute the Weierstrass $\wp$ functions associated to the domain and codomain. This requires solving a system of differential equation, and the algorithms for that in turn require computing the truncated reciprocal and exponential functions of polynomials. We first demonstrate the algorithms for solving the differential equations, and then show how to compute the reciprocal and exponential functions efficiently enough to give these algorithms $O(M(n))$ complexity. For now, to analyze the complexity of solving the system of linear equations we will assume this complexity.

Stark's algorithm to recover the kernel polynomial of an isogeny requires solving a first order nonlinear differential equation. To show how to do that, we will first present an algorithm for solving first order linear differential equations, and then show how this can be used to solve the desired system of linear equations.

## B.1 Solving a system of first order linear differential equations

To solve a system of linear differential equations, we use the following algorithm from [2] 2.3, originating from [3].

**Input:** A degree $n$, univariate polynomials $a$, $b$, and $c$ in $K[z]$ of degree at most $n$, where $a(0) \neq 0$, and a scalar $\alpha$ in $K$.
**Output:** A polynomial $f$ such that

$$af' + bf = c \mod z^n$$

and $f(0) = \alpha$.

1. Let $B = b/a \mod z^{n-1}$.
2. Let $C = c/a \mod z^{n-1}$.
3. Let $J = \exp_n \left( \int CB \right)$.
4. Return
$$f = \frac{1}{J} \int CJ \mod z^n.$$

The correctness of this algorithm can be seen directly by verifying that $f$ does in fact satisfy the desired equations. Here $\int$ denotes the antiderivative. So we note that to calculate the antiderivative we require that $1, \cdots, n-1$ are units in $K$ (thus for positive characteristic $n \leq p$.) In steps 1, 2 and 4, we calculate reciprocals and hence have complexity $O(M(n))$. Computing the truncated exponential in step 3 has complexity $O(M(n))$ as well. Computing the antiderivatives in steps 3 and 4 are $O(n)$. Hence the whole algorithm is $O(M(n))$.

## B.2 Solving a system of first order nonlinear differential equations

Now to show how to solve the nonlinear system of differential equations we re-state a special case of an algorithm from [3] as relayed in [2] section 2.4.

**Input:** A polynomial $G$ in $K[t]$, scalars $\alpha$ and $\beta$ in $K$, and degree $n$.
**Output:** The polynomial $f$ such that $f'(z)^2 = G(f)(z) \mod z^n$ and $f(0) = \alpha$ and $f'(0) = \beta$ (here $G(f)$ indicates the polynomial formed by composing $G$ with $f$.)

1. Set $f = \alpha + \beta z$ and $s = 2$.
2. While $s < n$ do
   (a) Set $a = 2f'$.
   (b) Set $b = G'(f)$ (Where $G'$ denotes the derivative of $G$ with respect to $t$.
   (c) Set $c = G(f) - (f')^2$.
   (d) Use the algorithm for first order linear differential equations to solve for $f \mod z^s$ by computing $f_2$ such that
   $$af_2' + bf_2 = c$$
   with $f_2(0) = 0$.
   (e) Set $f = f_2 + f \mod z^s$.
   (f) Let $s = 2s - 1$.
3. Return $f$.

The correctness of this algorithm follows from the fact that if $f_1 = f \mod z^s$ then $f_2 = f - f_1 \mod z^{2s-1}$ is a solution of the linearized differential equation

$$2f_1'f_2' - G'(f_2) = G(f_1) - (f_1')^2.$$

Note that for this algorithm to work $1, \cdots, n-1$ must be units (i.e. $n \leq p$ in positive characteristic,) so that we can use the algorithm for solving first order linear differential equations. We briefly analyze the complexity here. Calculating $a$, $b$, and $c$ to precision $s$ requires $O(M(n))$ and, as argued above, solving the system of linear differential takes $O(M(s))$ as well. Thus equation (A.1) implies that the whole complexity is $O(M(n))$, as the magnitude of $s$ roughly doubles with each iteration.

## B.3 Polynomial reciprocal and exponential functions

Next, we describe the polynomial functions of truncated reciprocal and truncated exponential. Then we describe how to compute these values in time $O(M(n))$.

For the truncated reciprocal of degree $n$, one can guess that this means that given a polynomial $f$, the reciprocal polynomial $g$ is the polynomial such that

$$f(z)g(z) \equiv 1 \mod z^n,$$

however to compute this, we can apply the iterative formula:

$$g_i = -\frac{1}{f_0} \sum_{j=1}^{i} f_i g_{i-j}$$

for $i \geq 1$, where $g_0 = 1/f_0$, the reciprocal of the constant coefficient which must be nonzero. It is less clear what the truncated exponential, denoted $\exp_n(f)$, of a polynomial is. However, it is just the evaluation of the power series

$$\sum_{i=0}^{n-1} \frac{f^i}{n!} \mod z^n.$$

In each of these cases, assuming that the input polynomial $f$ is of degree $n$, using these straight forward iterative formulas requires $O(n^2)$ operations in $K$.

In the case of the algorithms for computing the $\wp$-function, using $O(n^2)$ algorithms for exponential and reciprocal this would dominate the complexity, leading to $O(n^3)$ algorithms. As the complexity of these algorithms is the bottleneck, it is prudent to investigate different algorithms. It turns out that there are algorithms for both reciprocal and exponential that have complexity $O(M(n))$. Both of these algorithms use a technique called Newton iteration. The Newton iteration approach is a generalization of Newton's method for finding roots. Whereas Newton's method finds roots by approximation in the usual Euclidean metric of analysis, Newton iteration uses the $p$-adic metric where $p$ is some prime ideal ([8] chapter 9.) Here we do not give a proof of correctness, rather we just state the iterations.

The Newton iteration for computing the reciprocal of $f$ is

$$g_{i+1} = g_i \left(2 - f h_i\right) \mod z^{2^{i+1}}$$

for $i \geq 0$ where $g_0$ is $1/f_0$. So computing iteration $i$ requires $O(M(2^{i+1}))$ operations in $K$. Thus equation (A.1) implies that computing the truncated reciprocal to precision $n$ requires $O(M(n))$ operations ([2] 2.1, [8] algorithm 9.3.)

The Newton iteration for computing the exponential of $f$ is

$$g_{i+1} = g_i \left(1 + f - \log_{2^{i+1}}(g)\right) \mod z^{2^{i+1}}$$

for $i \geq 0$, where $g_0 = 1$. Similarly to how we defined the exponential function on polynomials by the power series expansion, we can define the truncated logarithm as

$$\log_n(g) = -\sum_{i=1}^{n-1} \frac{1}{i}(1-g)^i \mod z^n.$$

However, the logarithm can also be obtained by computing the truncated power series of $g'/g$ and taking the antiderivative. The derivative and antiderivative operations on a polynomial take $O(n)$ operations, and, as we just saw, calculating the reciprocal takes $O(M(n))$ operations, so that calculating the logarithm of a polynomial can be computed in time $O(M(n))$. Thus it follows that iteration $i$ takes $O(M(2^{i+1}))$ operations in $K$. And again, by equation (A.1) it follows that computing the truncated exponential to precision $n$ requires $O(M(n))$ operations in $K$ ([2] 2.2.)

# Part V
# Elliptic curve isogenies in Sage

As of release 4.0.2, Sage [19] includes an implementation of elliptic curve isogenies. This implementation was written by the author as part of the research for this project. The purpose of this appendix is to briefly describe and advertise this new elliptic curve isogeny functionality in Sage.
First we initialize an elliptic curve:

```
sage: F = GF(19);
sage: E = EllipticCurve(F, [0,0,0,1,2]); E
Elliptic Curve defined by y^2 = x^3 + x + 2 over Finite Field of size 19
sage: E.order()
12
```

The order 3 subgroup of the points defined over $\mathbb{F}_{19}$ is $\{\infty, (8,3), (8,16)\}$.

Then, like in Vélu's formulas we can specify the isogeny by giving this list of points to the constructor:

```
sage: P = E((8,3))
sage: phi = EllipticCurveIsogeny(E, [0*P, P, 2*P]); phi
Isogeny of degree 3 from Elliptic Curve defined by y^2 = x^3 + x + 2
over Finite Field of size 19 to Elliptic Curve defined by
y^2 = x^3 + 9*x + 3 over Finite Field of size 19
```

Alternately, we can use the kernel polynomial $\psi(x) = x - 8$ to construct the isogeny as in the algorithms of sections 3.1.2 and 3.2:

```
sage: R.<x> = F[]
sage: phi = EllipticCurveIsogeny(E, x-8); phi
Isogeny of degree 3 from Elliptic Curve defined by y^2 = x^3 + x + 2
over Finite Field of size 19 to Elliptic Curve defined by
y^2 = x^3 + 9*x + 3 over Finite Field of size 19
```

An isogeny object can be called as a function to evaluate the result at points on the domain curve:

```
sage: P = E.random_point(); P
(14 : 9 : 1)
sage: phi(P)
(16 : 14 : 1)
sage: P = E.random_point(); P
(8 : 3 : 1)
sage: phi(P)
(0 : 1 : 0)
```

The rational_maps function returns the coordinate maps:

```
sage: phi.rational_maps()
((x^3 + 3*x^2 - 6*x + 7)/(x^2 + 3*x + 7),
 (x^3*y - 5*x^2*y - 4*x*y - 4*y)/(x^3 - 5*x^2 + 2*x + 1))
```

The codomain function returns the codomain of the isogeny:

```
sage: E2 = phi.codomain(); E2
Elliptic Curve defined by y^2 = x^3 + 9*x + 3
over Finite Field of size 19
```

The constructor can also work to generate the isogeny from the domain and codomain, and the equals operator has been overloaded so that it works with isogenies (even when they are instantiated in different ways):

```
sage: psi = EllipticCurveIsogeny(E, None, E2, 3)
sage: psi == phi
True
```

The dual function returns the dual isogeny:

```
sage: phihat = phi.dual(); phihat
Isogeny of degree 3 from Elliptic Curve defined by y^2 = x^3 + 9*x + 3
over Finite Field of size 19 to Elliptic Curve defined by
y^2 = x^3 + x + 2 over Finite Field of size 19
sage: P = E.random_point(); P
(17 : 7 : 1)
sage: phihat(phi(P)) == 3*P
True
```

For more complete and in-depth documentation of the sage EllipticCurveIsogeny class, see the Sage documentation.